\documentclass[twocolumn,times,a4paper]{aastex61}
\usepackage{amsmath,amstext}
\usepackage[figure,figure*]{hypcap}
\usepackage{newtxmath} 
\usepackage{enumerate}
\usepackage{xspace} 
\usepackage[caption=false]{subfig}

\makeatletter

\newcommand{\pow}[1]{\ifmmode{}^{#1}\else ${}^{#1}$\fi}
\newcommand{\HI}{{\text{H\MakeUppercase{\romannumeral 1}}}\xspace}
\newcommand{\HII}{{\text{H\MakeUppercase{\romannumeral 2}}}\xspace}
\newcommand{\Lya}{\ifmmode{\mathrm{Ly}\alpha}\else Ly$\alpha$\xspace\fi}
\newcommand{\Ha}{\ifmmode{\mathrm{H}\alpha}\else H$\alpha$\xspace\fi}

\newcommand{\cm}{\,\ifmmode{{\rm cm}}\else cm\fi}

\newcommand{\ergps}{\,{\rm erg}\,{\rm s}\ifmmode{}^{-1}\else ${}^{-1}$\fi}
\newcommand{\Mpch}{\,{\rm Mpc}\,\ifmmode h^{-1}\else $h^{-1}$\fi}
\newcommand{\snru}{\,\ifmmode{\mathrm{Myr}^{-1}}\else Myr${}^{-1}$\fi}
\newcommand{\kms}{\,\ifmmode{\mathrm{km}\,\mathrm{s}^{-1}}\else km\,s${}^{-1}$\fi}

\newcommand{\fc}{\relax\ifmmode{f_{\mathrm{c}}}\else  $f_{\mathrm{c}}$\xspace\fi}
\newcommand{\fccrit}{\relax\ifmmode f_{\mathrm{c,\,crit}}\else  $f_{\mathrm{c,\,crit}}$\xspace\fi}

\newcommand{\diff}{\ensuremath{\; \text{d}}}

\newcommand{\abb}[2]{{#1}{\small \uppercase\expandafter{\romannumeral #2}}}
\newcommand{\abbm}[2]{\ensuremath{{\rm #1}{\small\rm\scriptstyle \uppercase\expandafter{\romannumeral #2}}}}

\makeatother

\received{2017 September 13}
\revised{2018 February 27}
\accepted{2018 March 7}
\submitjournal{ApJ}

\begin{document}
\title{Unlocking the Full Potential of Extragalactic Ly$\alpha$ through Its Polarization Properties}
\shorttitle{Unlocking the Full Potential of Extragalactic \Lya through Its Polarization Properties}
\shortauthors{Eide, Gronke, Dijkstra and Hayes}

\correspondingauthor{Marius Berge Eide}
\email{eide@MPA-Garching.MPG.DE}

\author[0000-0002-4234-2012]{Marius B. Eide}
\affil{Max-Planck-Institut f\"ur Astrophysik, Karl-Schwarzschild-Stra\ss e 1, D-85741 Garching, Germany}
\author[0000-0003-2491-060X]{Max Gronke}
\affil{Institute of Theoretical Astrophysics, University of Oslo, Postboks 1029, NO-0315 Oslo, Norway}
\affil{Department of Physics, University of California, Santa Barbara, CA 93106, USA}
\author{Mark Dijkstra}
\affil{Institute of Theoretical Astrophysics, University of Oslo, Postboks 1029, NO-0315 Oslo, Norway}
\author[0000-0001-8587-218X]{Matthew Hayes}
\affil{Department of Astronomy, Oskar Klein Centre, Stockholm University, AlbaNova University Centre, SE-106 91 Stockholm, Sweden}

\begin{abstract}
    Lyman-$\alpha$ (Ly$\alpha$) is a powerful astrophysical probe. Not only is it ubiquitous at high redshifts, it is also a resonant line, making
Ly$\alpha$ photons scatter. This scattering process depends on the physical conditions of the gas through which Ly$\alpha$
propagates, and these conditions are imprinted on observables such as the Ly$\alpha$ spectrum and its surface brightness
profile. In this work, we focus on a less-used observable capable of probing any scattering process: polarization.
We implement the density matrix formalism of polarization into the Monte Carlo radiative transfer code \texttt{tlac}.
This allows us to treat it as a quantum mechanical process where single photons develop and lose polarization from
scatterings in arbitrary gas geometries. We explore static and expanding ellipsoids, biconical outflows, and clumpy
multiphase media. We find that photons become increasingly polarized as they scatter and diffuse into the wings of
the line profiles, making scattered Ly$\alpha$ polarized in general. The degree and orientation of Ly$\alpha$ polarization
depends on the kinematics and distribution of the scattering HI gas. We find that it generally probes spatial or
velocity space asymmetries and aligns itself tangentially to the emission source. We show that the mentioned
observables, when studied separately, can leave similar signatures for different source models. We conclude by
revealing how a joint analysis of the Ly$\alpha$ spectra, surface brightness profiles, and polarization can break these
degeneracies and help us extract unique physical information on galaxies and their environments from their
strongest, most prominent emission line.
\end{abstract}

\keywords{galaxies: high-redshift -- galaxies: halo -- galaxies: ISM -- line: formation -- polarization -- radiative transfer}

\section{Introduction}
Lyman-$\alpha$ (\Lya) is the torch that lights up the distant Universe.
\cite{Partridge1967} recognized \Lya to be the strongest tracer of recombining ionized hydrogen (\HII) in young, (star) forming galaxies. However, the search for redshifted Ly$\alpha$ emission was fruitless until the late 1980s, when \Lya finally was found in known radio galaxies (see e.g.~\citealt{Djorgovski1985,Hu1987} or the overview by~\citealt{Spinrad1989}). Today, \Lya heralds the presence of the most distant sources known to humankind (e.g.~through
absorption, \citealt{Oesch2016} or emission, \citealt{Zitrin2015}), and detecting \Lya has become one of the primary science goals of future instruments and telescopes that are developed to understand the high-$z$ Universe (see e.g.~reviews by \citealt{Dijkstra2014} and \citealt{Hayes2015}).

The question remains what one can learn from observations of \Lya emission (and/or absorption). Thus far, observational efforts, as well as theoretical advances geared toward \Lya radiation, have focused primarily on the modulation of {\it intensity}. However \Lya radiation (or radiation of any wavelength) possesses two more degrees of freedom\footnote{A third degree of freedom also exist for circularly polarized light: the time-dependence of the polarization angle, expressed through Stokes $V$ parameter.}, which quantify its polarization properties. These are often represented through the Stokes $Q$ and $U$ parameters, and give the direction and degree of polarization. In this work we explore what additional knowledge can be obtained from these observables.

The potential power of \Lya lies in its resonance nature. In contrast to \Ha, which escapes unobstructed from its production site following recombination, a \Lya photon can undergo a tremendous number of scatterings after creation, where the precise number depends on HI column density, geometry and kinematics \citep[][]{Adams1972,Dijkstra2014}. Each scattering event results in a slight change in position and frequency. This dual diffusion process \citep{Osterbrock1962} imprints signatures on the emergent observables, and potentially reveals properties of the scattering medium along the paths that offered least resistance to the photons \citep[see e.g.~][]{Dijkstra2016,Gronke2016}.

These signatures can also act as keys to uncovering the emission mechanism. 
Centrally emitted \Lya photons, e.g.~when hey were created as nebular emission powered by Pop II stars \citep{Chapman2004}, Pop III stars \citep{Schaerer2002,Schaerer2003} or a nuclear black hole\footnote{Such spectrally hard sources would leave notably large \Lya equivalent widths.}  \citep{Geach2009} must scatter significantly in most cases prior to escape.
Spatially extended \Lya emission can be produced by inflowing, cooling gas \citep{Haiman2000}, gas that has been shock heated by supernova explosions \citep{Mori2004} or by galactic superwinds \citep{Taniguchi2000}, or as fluorescent radiation from an external ionizing field \citep{Hogan1987,Cantalupo2005}.  These photons do not need to escape from the dense interstellar medium (ISM), and therefore typically scatter less.
With \Lya ubiquitously present in galaxies, surveys provide a wealth of observations open for interpretation \citep[e.g.~][]{Steidel2011,Wisotzki2016,Shibuya2017,Herenz2017}. Theoretical work exploring the modulation of Ly$\alpha$ observables by radiative transfer effects, aim to convert these observables into constraints on the physical conditions of the gas in and around galaxies.

Currently, two quantities provide the main observables. One is the \textit{spectrum}, which encodes information on the frequency diffusion process of the photons, which leads to broadening and shifting of the spectral line shape \citep{Neufeld1990,Dijkstra2006}, by an amount which depends on kinematics, geometry and dust content of the scattering medium \citep[e.g.][]{Ahn1998,Hansen2006,Verhamme2006,Dijkstra2008,Gronke2015}.
These models been successful at reproducing observations \citep{Verhamme2008,Hashimoto2015,Karman2016,Yang2017a}, though it still unclear how physically realistic the models are (see e.g. \citealt{Gronke2016,Gronke2016a,Gronke2017}).

One problem is that widely different models can provide similar spectra. For example: the large majority of emission sources having a \Lya peak that is redshifted with respect to other lines in the system \citep{Kunth1998,Trainor2015}, something that can be easily explained by Ly$\alpha$ scattering through a galactic outflow \citep{Verhamme2006,Dijkstra2006}. However, it is known that the intergalactic medium (IGM) can also process away \Lya mainly in the blue part of the intensity spectrum, which can leave an intrinsically symmetric emission line with a net redshift \citep{Dijkstra2007,Laursen2011}. \cite{Dijkstra2008} showed that these different models give rise to different levels of polarization. This illustrates that polarization, when combined with spectroscopy, may tell models apart that otherwise are indistinguishable.

The other quantity, the surface brightness profile, can reveal the spatial diffusion process that Ly$\alpha$ photons undergo, before escaping possibly far from the site of emission. The resulting \Lya nebulae have been detected around many \Lya emitting galaxies \citep{Hayes2013,Wisotzki2016}, with larger counterparts around many quasars \citep{Cantalupo2014,Lake2015,Hennawi2015,Cai2017}, but not all \citep{Herenz2015}.

With integral field unit spectrometers (IFUs) such as MUSE \citep{Bacon2015}, or deep imaging surveys as SILVERRUSH \citep{Ouchi2017}, the number of detailed \Lya observations -- that is, spectra and sometimes surface-brightness information -- exceed thousands.
There are far fewer observations of polarized \Lya \citep{Prescott2011,Hayes2011,Humphrey2013,Beck2016,You2017}.
This is partially due to the observational difficulty associated with polarization measurements of distant sources. 
Polarization-equipped instruments have presently small fields of view (FOV), and multiplexed observations of the Stokes parameters is generally hard.
However, another reason is a lack of theoretical foundation which makes \Lya polarization results difficult to interpret.
This is something we wish to improve upon with this work.

Presently, there are two ways of implementing polarized \Lya transfer in numerical codes. The {\it first} approach treats polarization solely in the macroscopical sense, and assumes that all photons are 100\% linearly polarized, by `carrying' a polarization vector in addition to its direction vector and frequency \citep{Angel1969,Rybicki1999,Dijkstra2008,Trebitsch2016}. The {\it second} approach is that of \citet{Lee1994} \citep[also used in][]{Lee1997a,Lee1997,Lee1998,Lee1999,AHN2015POLARIZATIONHYDROGEN,Chang2017}, who employ a quantum mechanically precise treatment of scattering and polarization using density matrices, allowing unpolarized photons to {\it develop} polarization through scatterings (and also allowing polarized photons to become {\it depolarized}). We employ this latter method, as it is quantum mechanically more accurate, and implement it in the Monte-Carlo radiative transfer code \texttt{tlac} \citep{Gronke2014}.

The goal of this paper is to explore what additional information is encoded within the polarization properties of \Lya on the physical properties of the scattering medium. More concretely, our goal is to make go beyond the `standard' predictions for intensity $I$, and focus on two linear polarization parameters $Q$ and $U$, and to see whether this extra information can break degeneracies between different models, and this to gain a deeper physical understanding of sources of \Lya.

This work is structured as follows: we describe the detailed numerical implementation of the density matrix formalism of Ly$\alpha$ polarization in \S~\ref{sec:MC_polarization}. This section is technical and can be skipped for readers who are mostly interested in the results, which we present in \S~\ref{sec:results}. We discuss our results in more detail and in a broader context in \S~\ref{sec:discussion}, before concluding in \S~\ref{sec:conclusion}.

\section{Lyman-$\alpha$ Monte-Carlo polarization}
\label{sec:MC_polarization}
Radiative transfer is the art of describing the complex and arduous journey light takes on after being emitted. The equation of radiative transfer\footnote{For instance, given by Eq.~(1) in the review by \cite{Dijkstra2014} in its differential form.} illustrates this: a change in intensity at one frequency $\nu$ along an differential path length is affected by three factors: (1) attenuation, (2) emission and (3) redistribution in both space and frequency. The third factor is of paramount importance for Ly$\alpha$. It expresses any contributions into the intensity that did not originate at the same frequency or from the same direction. It is thus an integral over all frequencies and all solid angles embedded in a differential equation.

Monte-Carlo methods are the preferred way of treating radiative processes where the photons do not alter the state of the medium they travel through, but still are sensitive to the redistributions caused by scattering through it  \citep[see eg.~][]{Avery1968,Lee1997,Loeb1999,Ahn2000,Zheng2002,Dijkstra2008,Pierleoni2009,Laursen2010} because -- albeit slow -- they guarantee convergence even in complex density or velocity fields.

Here, we describe the basics of polarization in \S~\ref{sec:MC_emission}-\S~\ref{sec:MC_detection}, and how we implement the density matrix formalism into the radiative transfer code \texttt{tlac} \citep{Gronke2014} in \S~\ref{sec:MC_summary}.

\begin{figure}[tb]
    \begin{minipage}[c]{0.40\columnwidth}
        \centering
            \includegraphics[width=\columnwidth]{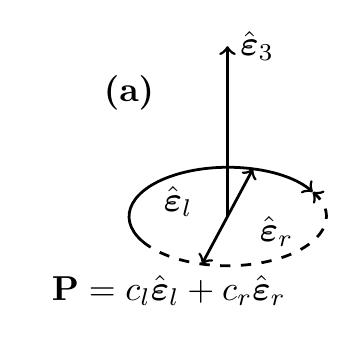}
            \hspace{2em}
    \end{minipage}
    \begin{minipage}[c]{0.48\columnwidth}
        \centering
            \includegraphics[width=\columnwidth]{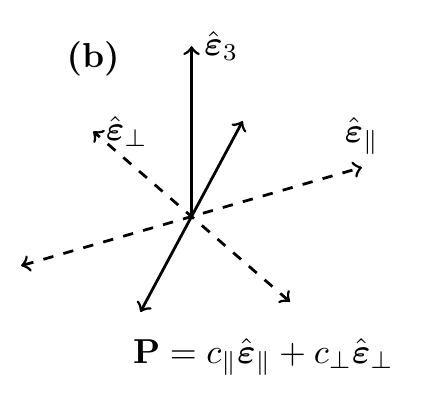}
    \end{minipage}
    \caption{Conceptual sketches of two possible choices of bases in the plane perpendicular to the photon propagation direction which is chosen to be $\hat{\boldsymbol{\varepsilon}}_3$: \textbf{(a)} complex, helical coordinates representative of the intrinsic spin of the photon, and \textbf{(b)} cartesian coordinates, representative of linear polarization appearing from a photon being in both spin states having a fixed, non-varying phase between the helical spins. The oscillations of the complex polarization vector $\mathbf{P}$ in the plane are also drawn.} 
    \label{fig:polarization_vector}
\end{figure}

\subsection{Emission}
\label{sec:MC_emission}
Ly$\alpha$ photons are emitted at, or near,
the Ly$\alpha$ resonance frequency of $\nu_0 = 2.47 \times 10^{15}$ Hz for hydrogen. We parametrize their offset from the line center through $x\equiv \left( \nu - \nu_0 \right)/\Delta \nu_{\rm D}$ where the Doppler width is defined as $\Delta \nu_{\rm D} = v_{\rm th} \nu_0/c$ with the thermal velocity $v_{\rm th} = \sqrt{2 k_{\rm B} T_g / m_p}$, which depends on the gas temperature $T_g$ as well as Boltzmann's constant $k_{\rm B}$ and the proton mass $m_p$. The speed of light is $c$.
We also parametrize the offset from the line center in terms of the thermal velocity.
The relationship is $v = - \lambda_0 \Delta \nu_{\rm D} x$. 
We also represent the spread in emission around the line center as the standard deviation of a Gaussian with $\sigma_i$ in units km s$^{-1}$.

The photons are massless and have four degrees of freedom through their two spins. Measuring the spin is synonymous to measuring their polarization. Intrinsically, the photons possess helical spins, whereas observationally, it is advantageous to consider the linear representation instead, and include a possible phase relation.

Following the approach of \cite{Lee1994}, we construct a complex state vector $\mathbf{P}$ with four degrees of freedom, represented through the complex coefficients $c_1$ and $c_2$, given in an orthogonal basis $\{\hat{\boldsymbol{\varepsilon}}_1, \hat{\boldsymbol{\varepsilon}}_2, \hat{\boldsymbol{\varepsilon}}_3 \}$ (where $\hat{\boldsymbol{\varepsilon}}_3$ denotes the propagation direction) as
\begin{equation}
    \mathbf{P} = c_1 \hat{\boldsymbol{\varepsilon}}_1 + c_2 \hat{\boldsymbol{\varepsilon}}_2.
    \label{eq:polarization_vector}
\end{equation} For a given $\mathbf{P}$, the values of the coefficients $c_1$ and $c_2$ depend
on the choice of basis, which is determined by what we desire to observe: circular or linear polarization.
For example, a helical basis is best suited to describe {\it circular polarization}, and we may use $c_1 \hat{\boldsymbol{\varepsilon}}_1 = c_l \hat{\boldsymbol{\varepsilon}}_l$ and $c_2 \hat{\boldsymbol{\varepsilon}}_2 = c_r \hat{\boldsymbol{\varepsilon}}_r$ to represent the left- and right-handed components of the spin, with probabilities $|c_l^2|$, $|c_r^2|$ of finding the photon being either left- or right-handed polarized, respectively. On the other hand, {\it linear polarization} arises from the superposition of the helical spins, describing it in terms of a parallel and perpendicular component, these can be written as $c_\parallel \hat{\boldsymbol{\varepsilon}}_\parallel =  ( c_l \hat{\boldsymbol{\varepsilon}}_l + c_r \hat{\boldsymbol{\varepsilon}}_r)/\sqrt{2}$ and $c_\perp \hat{\boldsymbol{\varepsilon}}_\perp = i(c_l \hat{\boldsymbol{\varepsilon}}_r - c_r \hat{\boldsymbol{\varepsilon}}_r)/\sqrt{2} $, respectively (see Figure~\ref{fig:polarization_vector} for a conceptual sketch of these representations of the polarization).

We may ask: if the square of the coefficients are equally large: is a polarization signal observable? The answer lies in the phase delay between the components: if both coefficients have equal magnitude {\it and} no fixed phase delay exists, there is no polarization. If there is a phase delay however, we will obtain a polarization signal. These additional constraints may be obtained from the cross-terms $c_1 c_2^*$ and $c_1^* c_2$. This discussion illustrates that the \textit{density matrix} of the photon, $\rho_{\rm phot}$, contains all information on its quantum state,
\begin{equation}
    \rho_{\rm phot} = \mathbf{P} \mathbf{P}^\dagger
    = 
    \left( {
        \begin{array}{cc}
            c_1 c_1^* & c_1 c_2^* \\
            c_2 c_1^* & c_2 c_2^*
        \end{array} }
    \right),
    \label{eq:dens_matrix}
\end{equation}
where the off-diagonal elements give the time-dependent phase between the two states, and the diagonal elements give the probabilities of measuring the photon in either of the two states.

\subsection{Scattering}
\label{sec:MC_scattering}
After emission, the photons may scatter with neutral hydrogen gas particles\footnote{In this work we focus exclusively on scattering by HI-atoms. Scattering by dust and electrons can be included in future studies. However, although dust clearly plays an important role in the \Lya radiative transfer process, its effect is mostly to destroy \Lya photons.}.
This interaction excites the particle from its ground state to an intermediate state, which it immediately de-excites into its final state. Should the initial and final states be the same, the photon will neither gain nor loose any energy\footnote{This is not entirely true. The photons deposit and gain energy through atomic recoil \citep{Madau1997} and hyperfine excitation of the ground state \citep{Wouthuysen1952,Field1958}}. However, for our usage cases these effects can be ignored \citep{1971ApJ...168..575A}. and the scattering is \textit{elastic}. We will not treat inelastic Raman scattering, but refer the interested reader to \cite{Lee1997} for an in-depth study of the polarization properties of Raman-scattered light. 

When a Ly$\alpha$ photon elastically scatters it experiences three types of redistributions: (i) \textit{change in propagation direction}, (ii) \textit{change of frequency}, and ({\it iii}) the change of polarization. We discuss each below.

\begin{itemize}
\item {\bf Change of propagation direction.} The change of direction is quantified by the {\it phase-function}, which we denote with $p(\theta', \phi' |\, \rho_{\rm phot}, \theta, \phi)$. Primed quantities denote scattered values. As we describe below, this phase-function depends on frequency and polarization of a Ly$\alpha$ photon.
It gives the probability of a photon being in the state it would obtain following a scattering, and can be directly related to the density matrix;
\begin{equation}
    p(\theta', \phi' |\, \rho_{\rm phot}, \theta, \phi) = \frac{|c_1^{2'}| + |c_2^{2'}|}{\int |c_1^{2'}| + |c_2^{2'}| \diff \Omega}.
    \label{eq:phase_function_chandra}
\end{equation} 
Each scattered density matrix component is obtained by a linear combination of the three incoming components, with trigonometric functions weighing each contribution.
Expressions for the density matrix are complex (see e.g.~\citealt{Lee1994} or \citealt{Lee1994a} for prescriptions for obtaining them, or \citealt{AHN2015POLARIZATIONHYDROGEN} for the relevant expressions for \Lya), and we refer interested reader to these papers.
Full expressions for $c_1^{2'}$ and $c_2^{2'}$ as a function of $(\theta', \phi',\rho_{\rm phot}, \theta,\phi)$ are given in Appendix~\ref{app:density_matrices}.

\item {\bf Change of frequency.} The `type' of elastic scattering depends on the offset from the resonance frequency $\nu_0$, as seen from the scattering atom. We express the velocity $v_{\rm atom}$ of the scattering atom as a dimensionless velocity $\mathbf{u} = \mathbf{v}_{\rm atom}/v_{\rm th}$. The frequency shift of the Ly$\alpha$ photon in the rest frame of the atom, $x_e$, is then
\begin{equation}
    x_e = x_i - \mathbf{u} \cdot \hat{\mathbf{k}}
    \label{eq:x_excitation_frequency}
\end{equation}
where $x_i$ was the initial frequency shift \citep{Laursen2010}. We can differentiate between \textit{resonance scattering} ($x_e \sim 0$) and \textit{wing scattering} ($|x_e| \gg 0$). This distinction is important: \cite{Stenflo1980} showed that, for resonant scattering the polarization properties of scattered Ly$\alpha$ relate to the \textit{spin} properties of the atomic configuration of the H-atom. On the other hand, for wing scattering, the electron behaves as if it is free. 

The transition from core to wing occurs at a temperature-dependent frequency offset $x_{cw}\sim 3$ \cite[see e.g.][for an expression for $x_{cw}$]{Laursen2010}. We also use an acceleration scheme for Ly$\alpha$ Monte-Carlo radiative transfer as in \cite{Dijkstra2006}, but have explicitly verified that our results are not affected by this.
\item {\bf Change of polarization.} The change of polarization properties is quantified by the change of the density matrix per scattering event.
    The newly obtained total\footnote{In the absence of circular polarization, $c_1 c_2^* = c_1^* c_2$ if we have chosen a linear basis. This cross term then only gives the $U$ polarization. Otherwise it also gives the $V$ polarization.} degree of polarization of a photon $P_{\rm phot}$ following a scattering is the fraction of the linearly ($Q$ and $U$) and circularly ($V$) polarized intensity to the total intensity ($I$),
    \begin{align}
        P_{\rm phot}&(\theta',\phi' | \rho_{\rm phot}, \theta, \phi)
        = \frac{\sqrt{Q^2 + U^2 + V^2}}{I} \label{eq:polarization_general} \\
        &= \frac{\sqrt{\left(|c_1^{2'}| - |c_2^{2'}|\right)^2 + 2^2(c_1 c_2^{*})^{'}(c_1^{*} c_2)^{'}} }{|c_1^{2'}| + |c_2^{2'}|}
        \label{eq:polarization_function}
    \end{align}
    following \cite{AHN2015POLARIZATIONHYDROGEN}.
\end{itemize}

We now turn to discuss resonant and wing scattering in more detail, as the distinction between the two plays an important role in the above processes.

\subsubsection{Resonance scattering}
\label{sec:MC_scattering_resonance}
For $|x_e| < x_{\rm cw}$, we will consider scatterings dominated by the transition from the ground energy state of hydrogen, denoted\footnote{We use the notation $nL_J$, $n$: energy level, $L=0,1,2,3,\dots$ denoted $S,P,D,F,\dots$ for the orbital angular momentum quantum number and $J=L+S$ where $S=\pm 1/2$ is the electron spin.} 
 $1S_{1/2}$ to the excited $n=2$ state, comprising the two available orbital configurations $2S_{1/2}$ or $2P_J$, where the $2P_J$ level is degenerate into $J=1/2$ and $J=3/2$, and back to the final $1S_{1/2}$ state.

This degenerate upper state with the similar angular configuration is also found in other atoms, but with larger frequency separations than $\Delta \nu = 1.1 \times 10^{10}$ Hz \citep{Brasken1998} obtained for hydrogen.
We will therefore adopt the terminology from those transitions: for Ca II, the transition from $J=1/2 \to J=3/2 \to J=1/2$ is denoted K (or D$_2$ for Na I), while  for the transition $J=1/2 \to J=1/2 \to J=1/2$, it is denoted H (or D$_1$ for Na I).

\textit{H scattering}: The wave function of the $2P_{1/2}$ state has no angular dependence, and when it de-excites, conservation of momentum may result in a photon traveling in any direction, with any perpendicular polarization vector.
Transitions through this state will yield a constant, angle-independent phase function, and zero polarization independent on any prior polarization,
\begin{equation}
p_{\rm H} (\theta', \phi' | \rho_{\rm phot}, \theta, \phi) = \text{const},
\label{eq:phase_K}
\end{equation}
with the subsequent density matrix being $|c_1^{2'}| = |c_2^{2'}| = 1/2$, and $c_1' c_2'^* = c_2' c_1'^* = 0$.

\textit{K scattering}: The wave function of the $2P_{3/2}$ state, on the other hand has a strong angular dependence.
The phase function now depends on all the density matrix coefficients, and hence also on the incoming polarization.
We present how the density matrix elements transform in Eqs.~(\ref{eq:K_11}--\ref{eq:K_22}), as given in Eq.~(11) in \cite{AHN2015POLARIZATIONHYDROGEN}\footnote{Or Eq.~(5) in \cite{Ahn2002}.}. 
These transformations are given for a left-handed photon basis, with one vector parallel to the plane of the scattering, and the other perpendicular to it.
The elements of the scattered density matrix obtained here are linear combinations of the elements of the incoming matrix, where the weights are determined by the incoming and scattered angles, as well as their differences.

\textit{Core scattering}:
We will from now refer to the resonant H- and K-transitions collectively as \textit{core scatterings.}
The small frequency separation between the two make it difficult to determine exactly the transition type.
However, we use that the effective ratio between the cross sections are $2\lambda_{\rm H}/\lambda_{\rm K} \approx 2$ \citep{Stenflo1980}.
In the resonance core, H-scattering is then twice as likely as K-scattering.

\subsubsection{Wing Scattering}
\label{sec:MC_scattering_wing}
As shown by \cite{Stenflo1980}, scattering far from the line center will, due to the interference between the two available sublevels of the excited Ly$\alpha$ state, resemble that of a classic oscillator. Wing scattering may be approached as a $J=0 \to J=1 \to J=0$ transition, which is the one representing Rayleigh and Thomson scattering alike \citep{Chandrasekhar1960}. For this transition, we obtain the phase function and degree of polarization from density matrix of Eq.~(4) in \cite{Ahn2002}. Scattering at right angles yields 100\% polarization, while light that is forward or backward scattered, retains its initial degree of polarization and the phase relation---thus preserving the polarization direction as well.

\subsection{Escape and Detection}
\label{sec:MC_detection}
Detection is the last step involved in the Monte Carlo procedure. Observationally, polarization properties of radiation are quantified by the Stokers parameters. To construct these parameters, we need extract these from the polarization properties of individual photons in our Monte-Carlo simulation (which are quantified by the density matrix/polarization state vector $\mathbf{P}$).

We achieve this by constructing a $3\times3$ `observable' density matrix $\rho_{\rm obs}$ which projects the complex polarization state vector $\mathbf{P}$ (i.e. the density matrix, see Eq~\ref{eq:dens_matrix}) onto the plane of the sky defined by the observer. First, we specify the direction along which we `observe' our model. Without loss of generality, we define this direction to correspond to the $+z$ direction, and thus assume that plane of the sky corresponds to the $xy$-plane. We then only select those photons that escape within a solid angle $\omega$ from the $+z$ direction, and calculate the Stokes parameters for each photon in this subset as follows:
\begin{align}
    I &= |c_x^2| + |c_y^2| \label{eq:stokes_I} \\
    Q &= |c_x^2| - |c_y^2| \label{eq:stokes_Q} \\
    U &= c_x c_y^* + c_x^* c_y = 2 c_x c_y^* = 2 c_x^* c_y\label{eq:stokes_U} \\
    V &= i \left( c_x c_y^* - c_x^* c_y \right) = 0 \label{eq:stokes_V}
\end{align}
where the coefficients $|c^2_x|$, $|c^2_y|$, $|c^2_z|$ and their phase relations $c_x c_y^*$, $c_x c_z^*$ and $c_y c_z^*$ and how these relates to the (intrinsic) density matrix of the photon, $\rho_{\rm phot}$ are given in Appendix~\ref{app:translating_to_observer_basis}.
The last equalities of Eqs.~(\ref{eq:stokes_U}) and (\ref{eq:stokes_V}) further indicate that we have no circular polarization as we neither have emission of circularly polarized Ly$\alpha$ or processes that induce it.

We may then proceed to create images of the binned Stokes components, either for all frequencies, or further bin the photons given their frequency.
We may then define the \textit{degree of polarization},
\begin{equation}
    P = \frac{\sqrt{Q^2 + U^2}}{I},
    \label{eq:degree_of_polarization}
\end{equation}
and the relevant \textit{polarization angle},
\begin{equation}
    \chi = \frac{1}{2} \arctan \left( \frac{U}{Q} \right),
    \label{eq:angle_of_polarization}
\end{equation}
in line with observational work \citep{Hayes2011}.
The degree of polarization and the polarization angle are thus derived quantities from the primarily binned Stokes parameters we calculated for each photon.
Note also that $I^2 \geq Q^2 + U^2$ \citep[see eg.~][]{Rybicki1979}, meaning that both $Q$ and $U$ may be zero when the intensity is not.

We have tested our implementation against known solutions. In Appendix~\ref{app:slab} we have tested our code against scattering of a plane parallel, semi-infinite slab known from \cite{Chandrasekhar1960} for which \cite{AHN2015POLARIZATIONHYDROGEN} also obtained results; against scattering in a Hubble-expanding cosmological volume known from \cite{Rybicki1999} in Appendix~\ref{app:hubble}; and against the expanding shell of \cite{Dijkstra2008} in Appendix~\ref{app:expanding_shell}.
The density matrix implementation in \texttt{tlac} yields equal results to those of \cite{AHN2015POLARIZATIONHYDROGEN}. 
Additionally, it reproduces the degree of polarization as well as the surface brightness profiles for the expanding IGM and outflowing shell, even though the results that were compared to were obtained with the approach of \cite{Angel1969}, ie.~with fully polarized photons.

\subsection{Monte Carlo Implementation Summary}
\label{sec:MC_summary}
We implement the density matrix formalism for polarization into \texttt{tlac} as follows:

\begin{enumerate}
    \item We assign a $2 \times 2$ (possibly complex) density matrix $\rho_{\rm phot}$ to each photon. We emit photons in a random direction $(\theta,\phi)$, and unpolarized. In practise this means that we assign a density matrix with $|c_1^2| = |c_2^2| = 1/2$ with no time-dependent correlation between them, i.e.~$c_1 c_2^* = c_2 c_1^* = 0$. 
    
    \item We generate an \HI optical depth $\tau$ from the distribution $\exp(-\tau)$, and convert $\tau$ into a physical distance $s$ the Ly$\alpha$ photon travels before it escapes, by solving the line integral $\tau = \int_0^s dr' n_{\rm HI}(r')\sigma_{\alpha}(\nu[r'])$ (see \citealt{Gronke2014} for a more extended description of the code).

 \item The new propagation direction after scattering depends on the phase-function, which depends on the density matrix, which depends in turn on the type of scattering event (H \textit{vs} K \textit{vs} wing), and the density matrix of the photon prior to scattering.

   The frequency of the photon determines whether the scattering occurs in the damping wing or in the core. For wing scattering the post-scattering density matrix is given by\footnote{See also Eq.~(4) in \cite{Ahn2002}.} Eqs.~(\ref{eq:wing_11}--\ref{eq:wing_22}) . For core scattering, we draw a random number $\mathcal{R} \sim \mathrm{Unif}[0,1)$. If $\mathcal{R} > 1/3$ the scattering is H-type and the photon is depolarized ($\rho_{\rm phot,00}' = \rho_{\rm phot, 11}' = 1/2$, other elements zero). Otherwise, the scattering is K-type and the post-scattering density matrix is given by\footnote{See also Eq.~(11) in \cite{AHN2015POLARIZATIONHYDROGEN}.} Eqs.~(\ref{eq:K_11}--\ref{eq:K_22}). We sample from these density matrices using the rejection method:
   we draw a random set of trial polar angles
   $\theta',\phi'$  uniformly from a sphere, and calculate the corresponding post-density density matrix $\rho_{\rm phot}'$,
\begin{equation}
    \rho_{\rm phot}' = f(\theta',\phi'\,|\, \rho_{\rm phot}, \theta, \phi),
    \label{eq:rho_intrinsic}
\end{equation} which in turn translates to the phase function, Eq.~(\ref{eq:phase_function_chandra}).
The phase function returns a number, which we compare to a randomly drawn number $\mathcal{R} \sim \text{Unif}[0,1)$. If $p(\theta', \phi' \,|\, \rho_{\rm phot}, \theta, \phi) \geq \mathcal{R}$, we accept the proposed scattering angles $\theta',\phi'$ as well as the scattered density matrix $\rho_{\rm phot}'$ and the photon moves on.

    \item To create observable, well-defined Stoke parameters, the density matrix $\rho_{\rm phot}$ can be transformed into a $3 \times 3$ density matrix $\rho_{\rm obs}$ using Eqs.~(\ref{eq:rho_observer}--\ref{eq:rho_cs}) that is relative to the observer, which is equivalent to observing the photons with a photon counting device that is fixed in space and no longer is oriented perpendicular to the propagation direction of the photon. This introduces six new density matrix coefficients $|c_x^2|, |c_y^2|, |c_z^2|, c_xc_y^* = c_x^* c_y, c_x c_z^* = c_x^* c_z, c_y c_z^* = c_y^* c_z$.
    \item For a chosen coordinate axis (which one observes nadir), there will be a plane spanned by the other two coordinate vectors. For this plane, we obtain well-defined Stokes parameters $I,Q,U$ and $V$ (the latter is zero) through Eqs.~(\ref{eq:stokes_I}--\ref{eq:stokes_V}).
    \item The Stokes parameters of each photon can be binned (by eg.~frequency, radial bins, spatial pixels) to create observables for the chosen plane. Multiple planes can be combined by assuming symmetries to increase the signal-to-noise ratio. We choose to observe photons escaping within a cone of $\cos 18^\circ = 0.95$ of the axis observed nadir, similar to \cite{Trebitsch2016}, who chose $15^\circ$ and did not find the choice to strongly affect the results.
\end{enumerate}


\begin{figure*}[t]
    \centering
    \includegraphics[width=0.23\textwidth]{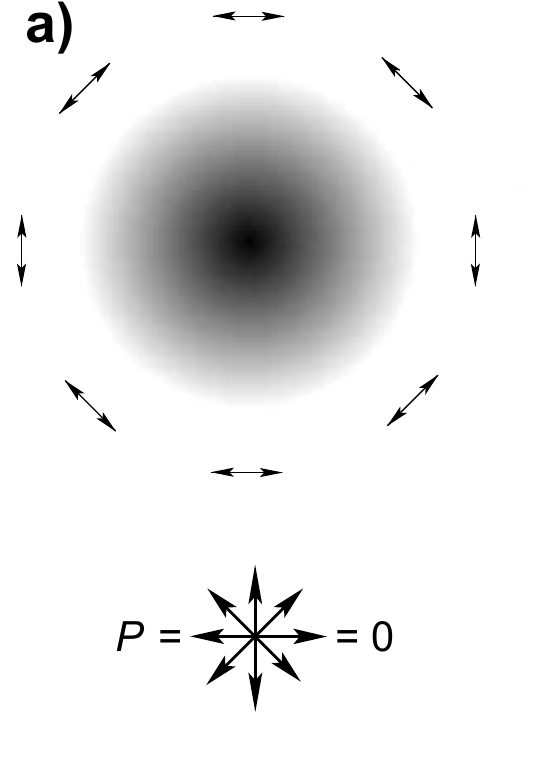}
    \includegraphics[width=0.23\textwidth]{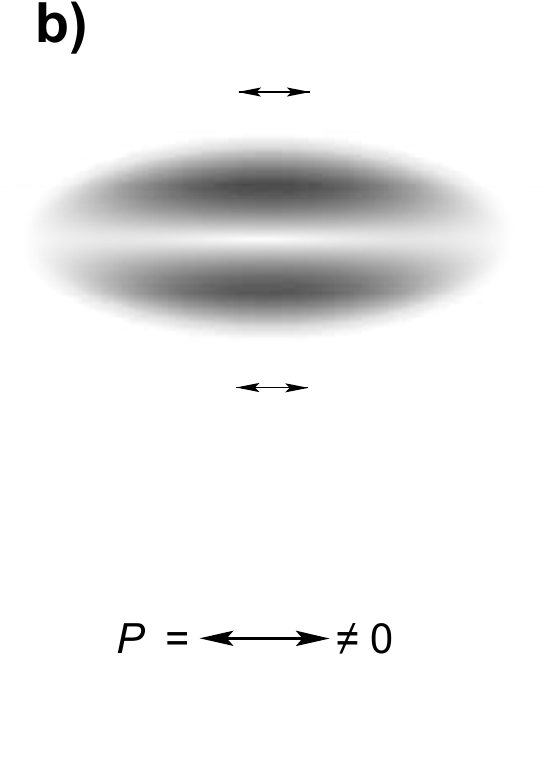}
    \includegraphics[width=0.23\textwidth]{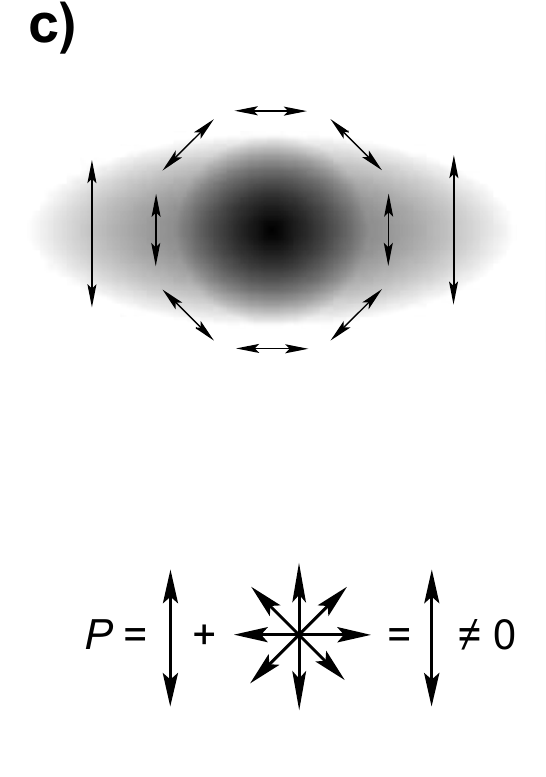}
    \includegraphics[width=0.23\textwidth]{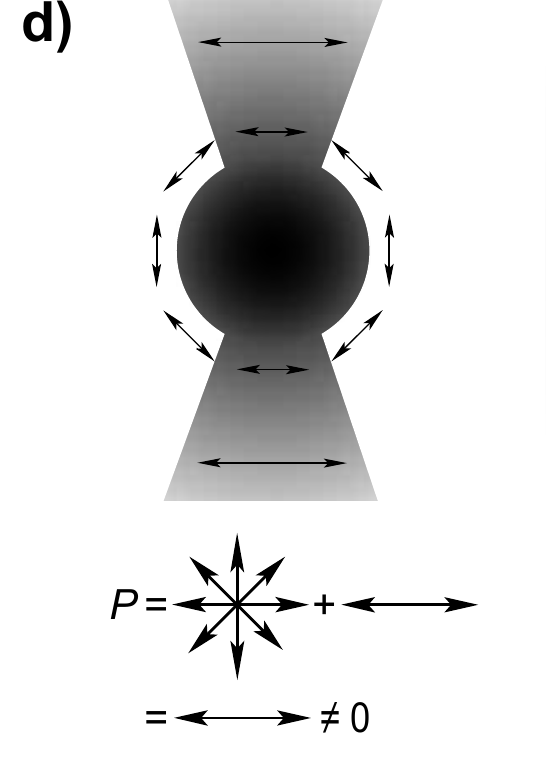}
    \caption{Sketches of four possible scattering geometries and their polarization signatures: \textbf{(a)} spherically symmetric scattering geometry where the polarization increases toward the limb and is tangential to it, \textbf{(b)} an oblate ellipsoid where the majority of the intensity is polarized parallel to the plane of the major axes, \textbf{(c)} an optically thinner ellipsoidal scattering geometry where the majority of the intensity is polarized perpendicular to the major axes and \textbf{(d)} a bipolar outflow where the polarization always is perpendicular to the outflow axis and a symmetrically polarized unobscured central core.
    \textit{In the lower panels}, we give the polarization if these extended sources were viewed edge-on as point sources, yielding \textbf{(a)} zero polarization as all vectors cancel due to the circular symmetry of the extended polarization signal, where photons that escape further out do so by scattering increasingly closer to $90^\circ$, leaving in sum a polarization signal that increases radially and that always is oriented tangentially to the central source; \textbf{(b)} non-zero polarization for an ellipsoid as the photons scatter and escape through optically thinner funnels along the minor axis of the ellipsoid before scattering at right angles toward the observer, becoming polarized horizontally; \textbf{(c)} non-zero polarization oriented perpendicular to the major axes of the ellipsoid as the scattering geometry was thin enough to allow photons to scatter along the major axes and then toward the observer, with the only allowed polarization being in the vertical direction; and \textbf{(d)} non-zero polarization oriented perpendicular to the outflow axis, as all the contributions from the brighter core are canceled out.  The shade indicates the intensity.}
    \label{fig:polarization_sketch}
\end{figure*}

\section{Results}
\label{sec:results}

\Lya radiative transfer through the interstellar and circumgalactic environments is a complex problem, and it is yet unclear which physical processes and scales play an important role in it.
It is therefore advantageous to study \Lya radiative transfer in simplified geometries in order to better identify the precise origins of the predicted observables, that is, in our case the predicted polarization signal. 
Here, we present calculations of Ly$\alpha$ polarization for a suite of simplified systems that have been adopted in the literature. These are representative of features in more complex astrophysical systems. 
In particular, we will discuss:
\begin{itemize}
\item static (\S~\ref{ssec:ellipsoid}) or expanding (\S~\ref{ssec:exp_ellipsoid}) ellipsoids,
\item biconical outflows (\S~\ref{ssec:outflows}), and
\item clumps of HI clouds, representative of a multiphase medium (\S~\ref{ssec:clouds}).
\end{itemize}

This can be thought of as a sequence in asymmetry: {\it first:} we introduce asymmetry in the gas distribution, {\it second:} we add an asymmetry in velocity space, {\it third:} we introduce further geometrical complexities by introducing biconical outflows, {\it fourth:} finally, we introduce `multiphase' versions of the outflow models. For each model, we will introduce the model parameters and present computed Ly$\alpha$ observables that could shed light on the nature and geometry of the sources and their environment. 
We sketch these models and some of our findings in Figure~\ref{fig:polarization_sketch}, which will be referred to throughout the text. Note that the apparent geometry of a system can change with frequency as \Lya photons of different frequencies escape at different spatial locations.

Following \cite{Lee1998}, we focus on computing the {\it frequency-dependence} of polarization for point sources, which differs from more recent analyses, which focused on (frequency-)integrated properties of spatially extended sources \citep{Dijkstra2008,Dijkstra2012a,Trebitsch2016}. This may represent a case where the Ly$\alpha$ source is spatially unresolved, or a case in which a spectroscopic slit is wide enough to cover the entire source.
For these point sources, we also show the total (i.e. integrated over frequency) polarization signal and its direction (relative to the unobservable geometry of the source in the plane of an observer).

\subsection{Oblate Ellipsoids}
\label{ssec:ellipsoid}

\begin{figure}[t]
    \centering
    \includegraphics[width=\columnwidth]{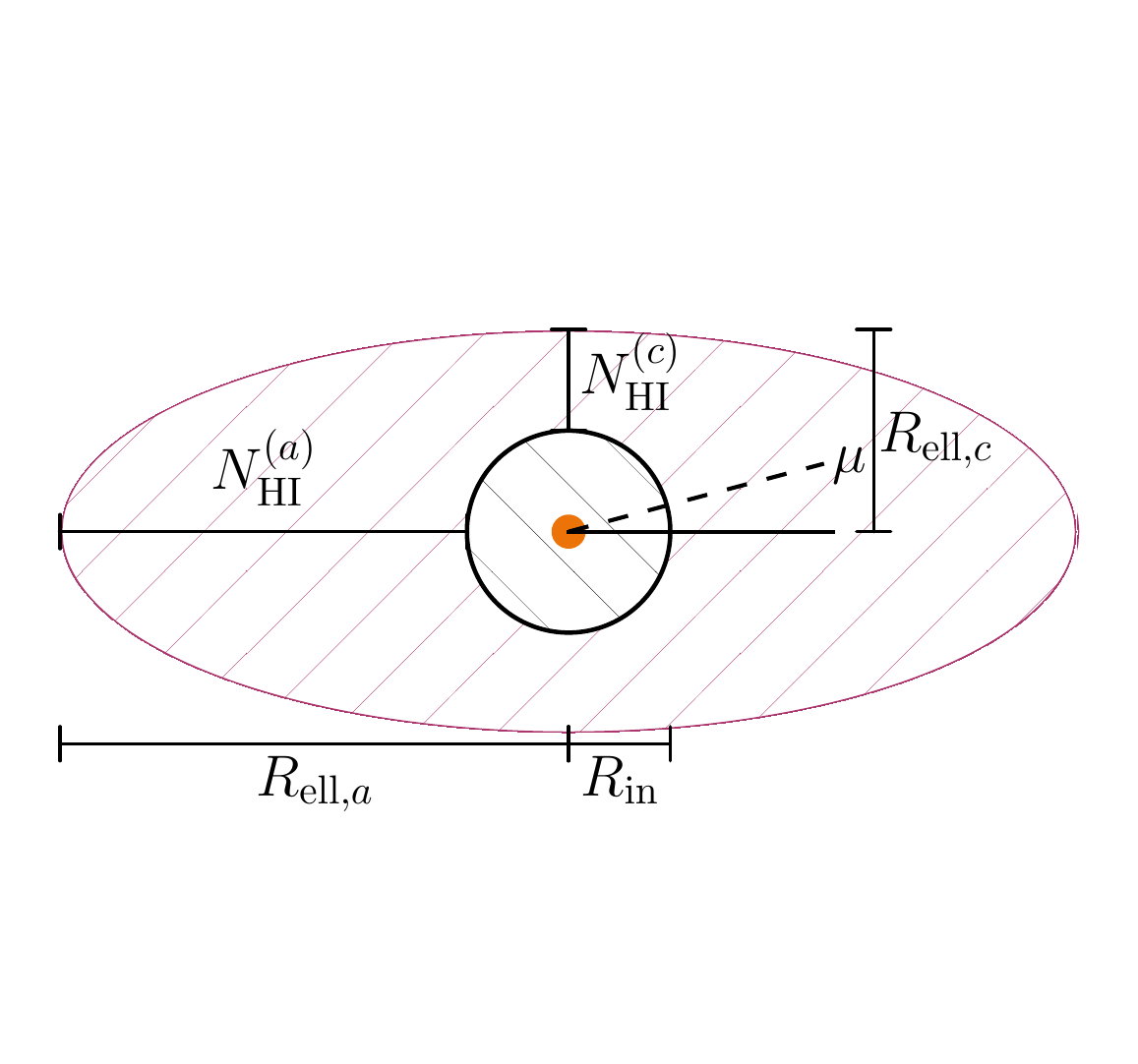}
    \caption{Ellipsoidal scattering geometry. 
        We have an ionized, inner region of radius $R_{\rm in}$ where our Ly$\alpha$ source is located, and an outer, ellipsoidal HI region with principal axes $\left( R_{{\rm ell},a}, R_{{\rm ell},b}, R_{ {\rm ell}, c} \right)$ where $R_{ {\rm ell}, a} = R_{ {\rm ell}, b}$. 
    The outer ellipsoid has neutral hydrogen column densities $N_{\rm HI}^{(a)}$ and $N_{\rm HI}^{(c)}$ along the principal axes. 
    The viewing angle is given as $\mu$.
}
    \label{fig:ellipsoid_geometry}
\end{figure}

\begin{figure*}[t]
    \subfloat
    [Intensity (normalized to unit area under curve) and polarization spectra as function of velocity offsets from the \Lya line center for ellipsoids
        with $N_\HI^{(c)} = 10^{21}$ cm$^{-2}$ along the minor axis, $T=10^4$ K, viewed edge-on as point sources.
    The colors indicate their ellipticities, given as the ratio between the minor and major axes, $\varepsilon \equiv R_{\rm ell,c}/R_{\rm ell,a}$. 
    The vertical dashed, grey line indicates a typical velocity offset where photons blueward commonly are seen to be absorbed by the IGM. 
    Hence, the spectrum blueward of the dashed line is in many cases not detectable.
    \label{fig:spectra_ellipsoids}
    ]
{\includegraphics[width=\columnwidth]{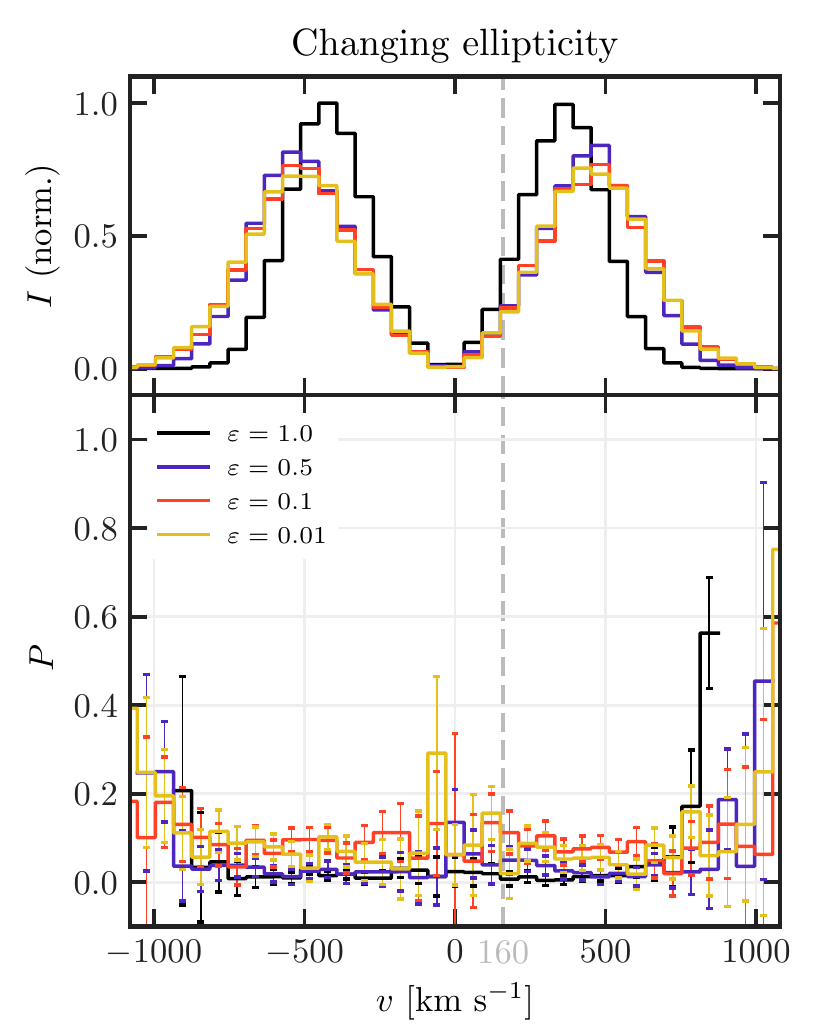}}
\qquad
    \subfloat
    [The degree and direction of polarization for ellipsoids with varying column densities along the minor axis, $N_{\rm HI}^{(c)}$ and varying ellipticities $\varepsilon$.
    The ellipsoids are viewed edge-on, ie.~where they appear to be asymmetrical, but as point sources to prevent introduction of non-intrinsic geometric asymmetries by for example a narrow slit viewing only parts of the overall source. 
    Viewed face-on, ie.~where the ellipsoids would have appeared symmetric to an observer, the total polarization is zero for all ellipticities and column densities, unlike in this plot.
    The direction of the arrows indicate the polarization direction relative to the plane of the major axes: horizontal arrows are parallel to it, whereas vertical arrows are perpendicular to it.
    \label{fig:matrix_polarization_ellipsoids}
]
{\includegraphics[width=\columnwidth]{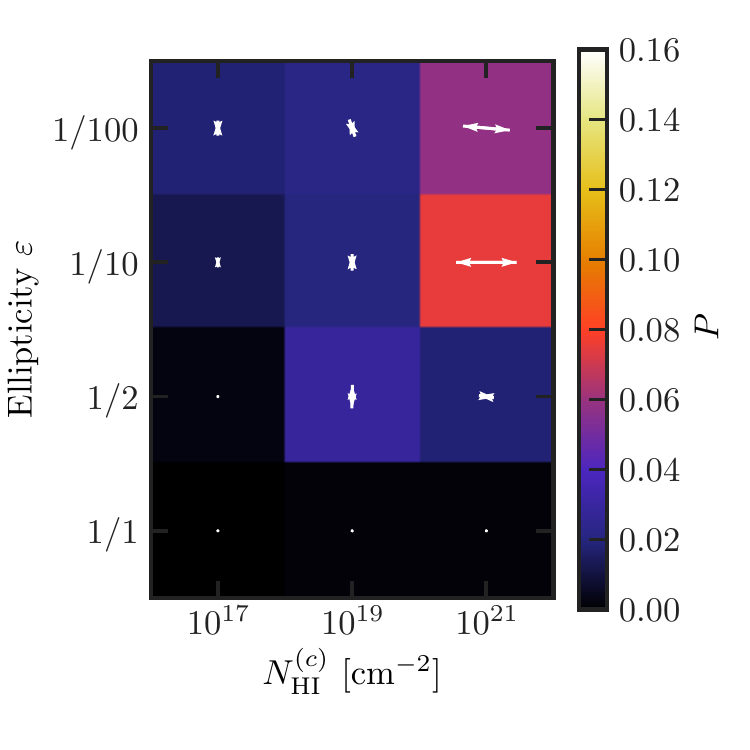}}
\caption{Static ellipsoids with varying ellipticity and column density along the minor axis, $N_\HI^{(c)}$.}
\end{figure*}

\cite{Angel1969} showed that Thomson scattering of thermal X-rays emanating from a ellipsoidal scattering geometry could provide up to 5\% polarization, if it was viewed from the side as a point source. This was further explored by \cite{Kim2007}, who considered H$\alpha$ line and continuum radiation emanating from an ellipsoid which also acted as a Rayleigh scatterer. They found a viewing angle-dependent increase in polarization in the wings.

   We pursue this idea further for Ly$\alpha$. We run a set of simulations where we explore the effects of \textbf{(i)} changing the column density of neutral HI gas, and \textbf{(ii)} changing the ellipticity. Changing the column density is known to have profound effect on the emergent spectrum. However, as shown in \cite{Dijkstra2016}, the shape of the spectrum emanating from a source region fully enclosed in an ellipsoid will primarily be given by the axis of lowest column density, and therefore it cannot reveal the ellipticity.

    We create an ellipsoid of static HI gas with fixed number density $n_{\rm HI}$ with principal axes $\left( R_{{\rm ell},a}, R_{{\rm ell},b}, R_{ {\rm ell}, c} \right)$ where we set the major axes $R_{ {\rm ell}, a} = R_{ {\rm ell}, b}$ and the perpendicular minor axis $R_{ {\rm ell}, c}$.
    This ellipsoid embeds an inner, ionized region with radius $R_{\rm in}$.
    In this inner region, we find our source of unpolarized Ly$\alpha$.
    The column densities along the principal axes can be found as $N_{\rm HI} = n_{\rm HI} (R_{\rm ell} - R_{\mathrm{in}})$, i.e., the neutral hydrogen number density is constant throughout the system.
    The viewing angle $\mu$ is defined relative to the plane of the two major axes (see Figure~\ref{fig:ellipsoid_geometry} for a sketch of this geometry).

    We choose three initial column densities along the minor axis, $N_{\rm HI}^{(c)}=10^{17}, 10^{19}, 10^{21}$ cm$^{-2}$. The choice of column densities reflect those expected in real systems \citep{Verhamme2017,Hashimoto2017,Gronke2015} 
    The lower bound, $N_\HI^{(c)} = 10^{17}$ cm$^{-2}$, corresponds to a case from which ionizing LyC may escape. The upper bound $N_\HI^{(c)} = 10^{21}$ cm$^{-2}$ corresponds roughly to the upper envelope of $N_{\rm HI}$ that is inferred from Ly$\alpha$ emitting galaxies. We then vary the ellipticity by varying the major axes (i.e. $R_{ {\rm ell}, a}$ and $R_{ {\rm ell}, b}$). This gives a set of ellipticities $\varepsilon \equiv R_{{\rm ell},c}/R_{{\rm ell},a}=\{1, 1/2, 1/10, 1/100\}$\footnote{A change in ellipticity is equivalent with changing the column density along the major axes, $N_{\rm HI}^{(a)}$.}. We fix $R_{\rm in} = 10$ pc and $R_{ {\rm ell},c} = 20$ pc and note that the choice of scale is arbitrary\footnote{However, the choice of the ratio $R_{\rm in}/R_{ {\rm ell},c}$ may not be, we have however not investigated this further.} for media that are static or have constant velocity fields. Our results are thus not scale-dependent. Furthermore, we set the gas temperate to $T=10^4\,$K and inject the photons with $\sigma_i = 200\kms$.

    In Figure~\ref{fig:spectra_ellipsoids}, we plot the spectra of the intensity $I$ and the polarization $P$ for the emergent photons for $N_\HI^{(c)}=10^{21}$ cm$^{-2}$ and all ellipticities, viewed with $\mu=0$ (edge-on).
    As a guide to the eye, we plot (in this and other intensity spectra) a gray dashed line, centered at 160 km s$^{-1}$. 
    This velocity offset marks a typical boundary between the redward, observable part of the spectrum, and the blueward, inaccessible part, for which the increasingly neutral IGM at higher $z$ prevents transmission of \Lya photons \citep{Dijkstra2007,Laursen2011}.
    
    As we are effectively changing the column density of the major axis, and the ($I$) spectra do not change, we realize that their shape are given by the column densities along the minor axis, which does not change, and not along the line of sight (LOS).
    In other words, the spectrum is mostly determined by the HI column density along the path of `least resistance' (see also \citealt{Dijkstra2016} for a similar result). However, we find that $P$ is overall higher across the spectrum for increased ellipticities, including at those frequencies where most photons escape.
    This can be understood as a consequence of the increasing deformation of the source with increasing ellipticity.
    At all frequencies, the shape of the source becomes asymmetric, leaving a preferential polarization direction and an overall non-zero polarization signal.
    
We quantify this effect in Figure~\ref{fig:matrix_polarization_ellipsoids}. Here we show the fractional degree of polarization $P$ as a function of \textbf{(i)} the minor axis column density $N_{\rm HI}^{(c)}$ and \textbf{(ii)} the ellipticity $\varepsilon$. We assumed that we view the sources edge-on\footnote{Face-on sources would appear circularly symmetric, and any polarization signal averages out.} (i.e. the projected size of the source is $R_{ {\rm ell},c} \times R_{ {\rm ell},a}$). The color of a bin indicates the degree of polarization. The arrow indicates the direction of the linear polarization w.r.t~the plane of the major axes (the size of the arrow also reflects its magnitude).

The {\it lower 3 panels} of Figure~\ref{fig:matrix_polarization_ellipsoids} show that the degree of polarization is $P<1$ \% for a spherical scattering geometry ($\varepsilon = 1$), as the spherical symmetry washes out any polarization. The polarization increases with ellipticity, but in a way that depends non-trivially on $N_{\rm HI}^{(c)}$: the direction of the polarization vector changes as $N_{\rm HI}^{(c)}$ increases from  $N_{\rm HI}^{(c)} = 10^{19}$ cm$^{-2}$ to $N_{\rm HI}^{(c)} = 10^{21}$ cm$^{-2}$. 
This change in column density effectively blocks all light from the central part of the system, leaving only photons that escape along the minor axes.
See panel b) of Figure~\ref{fig:polarization_sketch} for a sketch of this obscuration.
To reach the observer, they have to scatter closer to $90^\circ$, obtaining large degrees of polarization with the polarization vector oriented parallel to the major axes.
This effect is fundamentally similar to the effect seen in spherically-symmetric systems.
At large radii, photons also are highly polarized as they must escape these systems at $90^\circ$ to reach the observer, see e.g.~the rise in $P$ with radius in Figure~3 of \cite{Dijkstra2008}.
The polarization direction is always tangential to the central source.
In these systems, the global signal would be canceled out from symmetries, however, as illustrated in panel a) of Figure~\ref{fig:polarization_sketch}.

\subsection{Expanding Ellipsoids}
\label{ssec:exp_ellipsoid}

\begin{figure*}[t]
    \subfloat
    [Intensity and polarization spectra for ellipsoids with ellipticity $\varepsilon=1/10$, column density $N_{\rm HI}^{(c)} = 10^{19}$ cm$^{-2}$ along the minor axis, $T=10^4$ K, with varying degrees of global outflow velocities $v_{\rm flow}$ indicated by the line color. The expanding ellipsoids are viewed edge-on. The grey dashed line indicates the threshold for IGM removal of photons.
    \label{fig:spectra_expanding}
   ]
    {
    \includegraphics[width=\columnwidth]{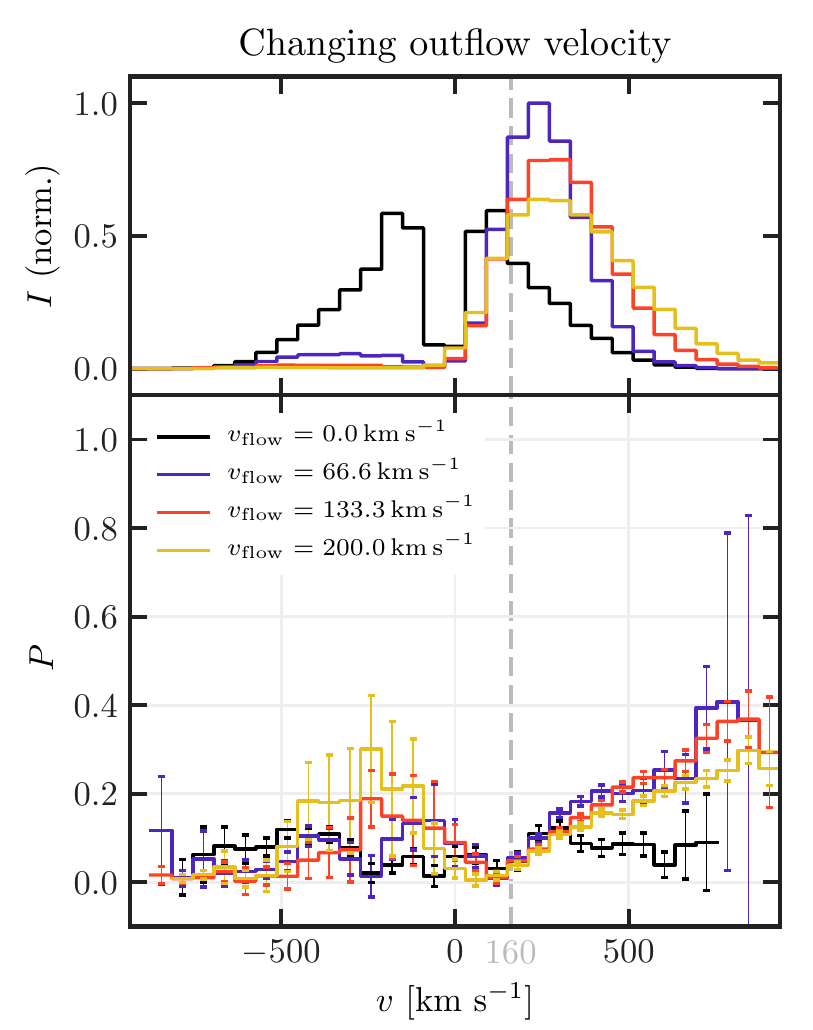}
    }
\qquad
    \subfloat
    [Total degree of polarization and its direction (given by the arrows) relative to the plane of major axes for globally expanding ellipsoids with outflow velocities $v_{\rm exp}$, ellipticity $\varepsilon=1/10$, and column densities along the minor axis $N_{\rm HI}^{(c)}$.
      The spectra of the sources with $N_{\rm HI}^{(c)} = 10^{19}$ cm$^{-2}$ are plotted in Figure~\ref{fig:spectra_expanding}.
    \label{fig:matrix_polarization_expanding_ellipsoids}
    ]
    {
    \includegraphics[width=\columnwidth]{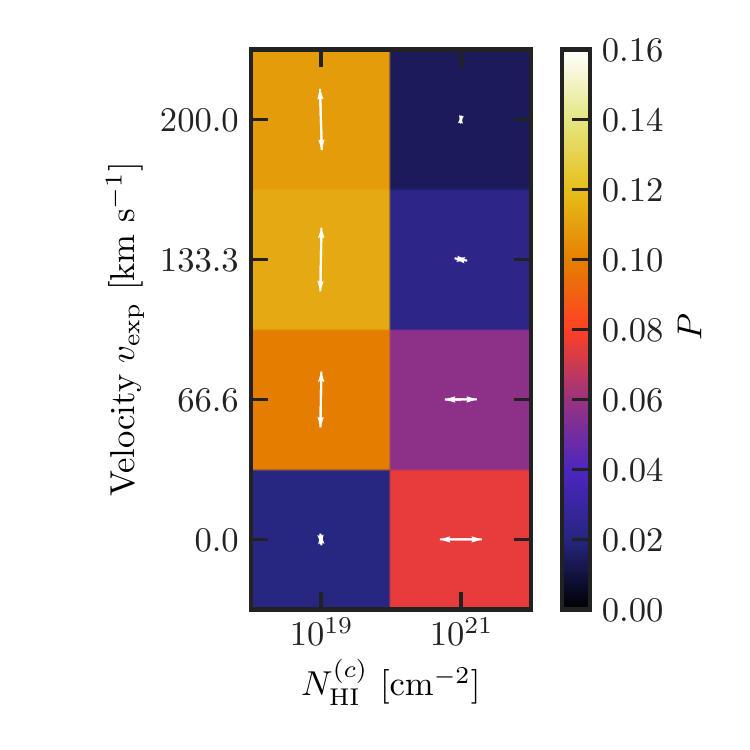}
    }
    \caption{Expanding ellipsoids with varying outflow velocity and column density along the minor axis, $N_\HI^{(c)}$.}
\end{figure*}

The previous section discussed the polarization emerging from static ellipsoids. Here, we add an outflowing component to the ellipsoid with constant velocities $v_{\rm flow} = \{0, 67, 200\}$ km s$^{-1}$, which is directed radially outward. As in \S~\ref{ssec:ellipsoid} we fix $\sigma_i = 200$ km s$^{-1}$. The presence of an outflow introduces an additional degree of asymmetry, now in velocity space.

Figure~\ref{fig:spectra_expanding} shows the spectrum and frequency-dependence of the polarization emerging from sources with fixed $N_{\rm HI}^{(c)} = 10^{19}$ cm$^{-2}$ along the minor axis, ellipticity $\varepsilon=1/10$, but with different expansion velocities $v_{\rm flow}$. For the static case we recover the double peaked spectrum that we obtained in \S~\ref{ssec:ellipsoid}. 
Expansion causes the majority of the photons to escape in the red wing, as blueward photons experience a higher optical depth \citep[see e.g.~][]{Zheng2002,Ahn2003,Dijkstra2006}.

The frequency dependence of the polarization is also asymmetric around the line center. For the cases with $v_{\rm flow} \neq 0$, there is little flux on the blue side of the line. The polarization of this flux is comparable to that of the static case (within the uncertainties). On the other hand, in the red wing of the line, outflows enhance the degree of linear polarization significantly. We obtain an increasing degree of linear polarization with velocity offset $\Delta v$ from the line center that approaches $P\sim 30\%$ asymptotically at $\Delta v > 500$ km s$^{-1}$. It is remarkable that the increase in the degree of linear polarization is very similar for all models with $v_{\rm flow} \neq 0$. 
This result can be understood as follows.
The distance a photon can travel increases for increasing outflow velocities, effectively lowering the optical depth seen by the photons.
In our cases, the change in outflow velocities would not necessarily imply a change in observed spatial shape of the system.
The optical depth is sufficiently low to make a significant fraction of the photons diffuse along the major axes in the presence of outflows, with polarization vectors tangential to the direction of the central source, as sketched in panel c) of Figure~\ref{fig:polarization_sketch}.
This means that the spatial asymmetry does not change significantly, and the degree of polarization remains similar between the models.

Figure~\ref{fig:matrix_polarization_expanding_ellipsoids} shows the degree and direction of polarization of expanding ellipsoids with $N_{\rm HI}^{(c)} = \{ 10^{19}, 10^{21} \}$ cm$^{-2}$ and $\varepsilon=1/10$ viewed edge-on. We omit the case $N_{\rm HI}^{(c)} = 10^{17}$ cm$^{-2}$, because we found that for this case, too few photons scatter, and our predictions practically correspond to that assumed for the intrinsic source. For $N_{\rm HI}^{(c)} = 10^{19}$ cm$^{-2}$, the polarization is near zero for the static ellipsoid (in agreement with the {\it central upper panel} of Figure~\ref{fig:matrix_polarization_ellipsoids}). When we increase $v_{\rm flow}$, the polarization remains roughly constant $P\sim10\%$, with the polarization direction aligned perpendicular to the major axes. 
This can be understood to be of the same reasons the degree of the polarization in the spectra did not change. 
The spatial (observed) shape of the systems do not significantly change for the increasing outflow velocities.
As the asymmetries do not change, the systems obtain similar degrees and directions of polarization.

Figure~\ref{fig:matrix_polarization_expanding_ellipsoids} contains other interesting results: For a higher column density along the minor axis, $N_{\rm HI}^{(c)} = 10^{21}$ cm$^{-2}$, the polarization behaves completely different compared to the case $N_{\rm HI}^{(c)} = 10^{19}$ cm$^{-2}$.  The total polarization of $P=8\%$ for a static ellipsoid (see also Figure~\ref{fig:matrix_polarization_ellipsoids}), and it now \textit{decreases} with outflow velocity, reaching $P=1\%$ for $v_{\rm flow} = 200$ km s$^{-1}$.
For this higher column density, the system is seen transitioning from the state sketched in panel b) in Figure~\ref{fig:polarization_sketch} to the more symmetric case presented in panel c).
However, there is one important difference from the system presented in panel c) as well as the system with lower column density $N_{\rm HI}^{(c)}$.
The increase in optical depth from the increased column density means that the photons also increasingly scatter once they encounter the rest-frame velocity offset of the atoms in the expanding medium.
This isotropizes the local radiation field, and the polarization vectors become randomized.
The lowering in optical depth from the increasing outflow means the photons penetrate deeper also along the major axes, contributing to removing the spatial asymmetry.
The overall effect is that the net polarization is reduced.
This reduction is eventually accompanied by a flip in the polarization vector of the $N_{\rm HI}^{(c)} = 10^{21}$ cm$^{-2}$ system, where the vector changes alignment from being parallel to perpendicular to the major axes. This indicates an increased semblance to the geometry of the much lower column density system $N_{\rm HI}^{(c)} = 10^{19}$ cm$^{-2}$.

\subsection{Bipolar Outflows}
\label{ssec:outflows}
\begin{figure}[t]
    \centering
    \includegraphics[width=0.49\columnwidth]{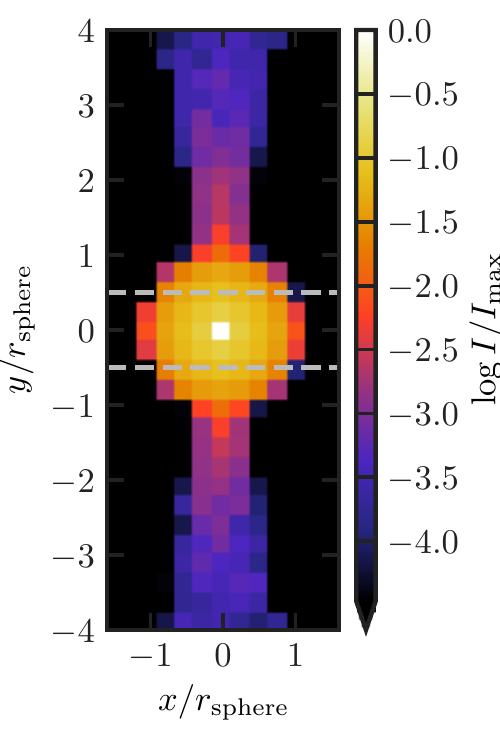}
    \includegraphics[width=0.49\columnwidth]{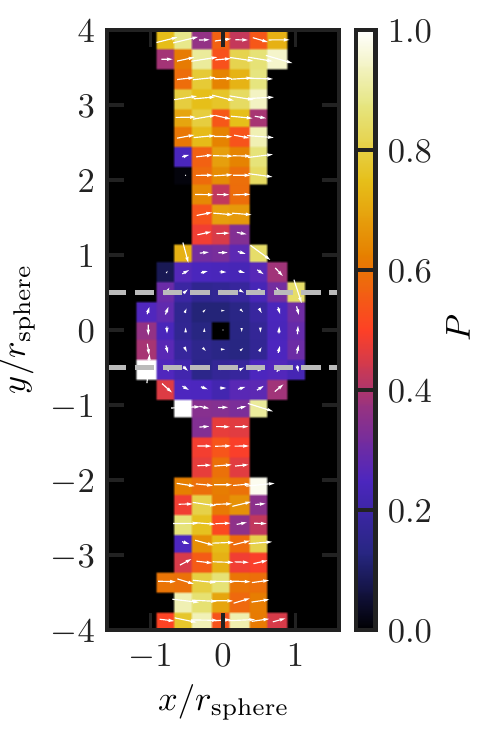}
    \caption{
        Spatially extended maps of the intensity and degree of polarization for bipolar outflows of $v_{\rm flow} = 200$ km s$^{-1}$ from a static sphere with $N_{\rm HI}=10^{19}$ cm$^{-2}$ and a total opening angle of $\theta_{\rm flow} = \pi/8$. 
    The color indicate the degree of polarization, and the arrows indicate the direction of polarization. 
    The dashed lines indicate the region that is obscured in Figure~\ref{fig:bipolar_spectrum}, to resemble the removal of Ly$\alpha$ photons by for example a dusty disk.
}
    \label{fig:bipolar_polarimage}
\end{figure}
\begin{figure}[t]
    \centering
    \includegraphics[width=\columnwidth]{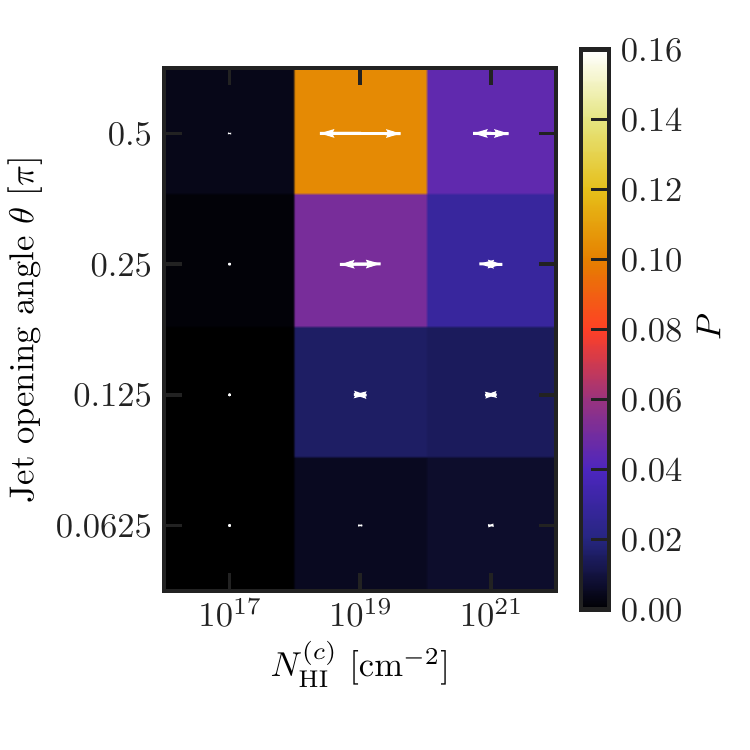}
    \caption{Degrees of polarization had the unobscured bipolar flows (with outflow velocity $v_{\rm flow} = 200$ km s$^{-1}$) been viewed as a point source. Direction of overlaid polarization vectors indicate their direction relative to the flow (horizontal: perpendicular to flow directions, vertical: parallel to flow directions).}
    \label{fig:bipolar_matrix}
\end{figure}

\begin{figure*}[t]
    \centering
    \includegraphics[width=\columnwidth]{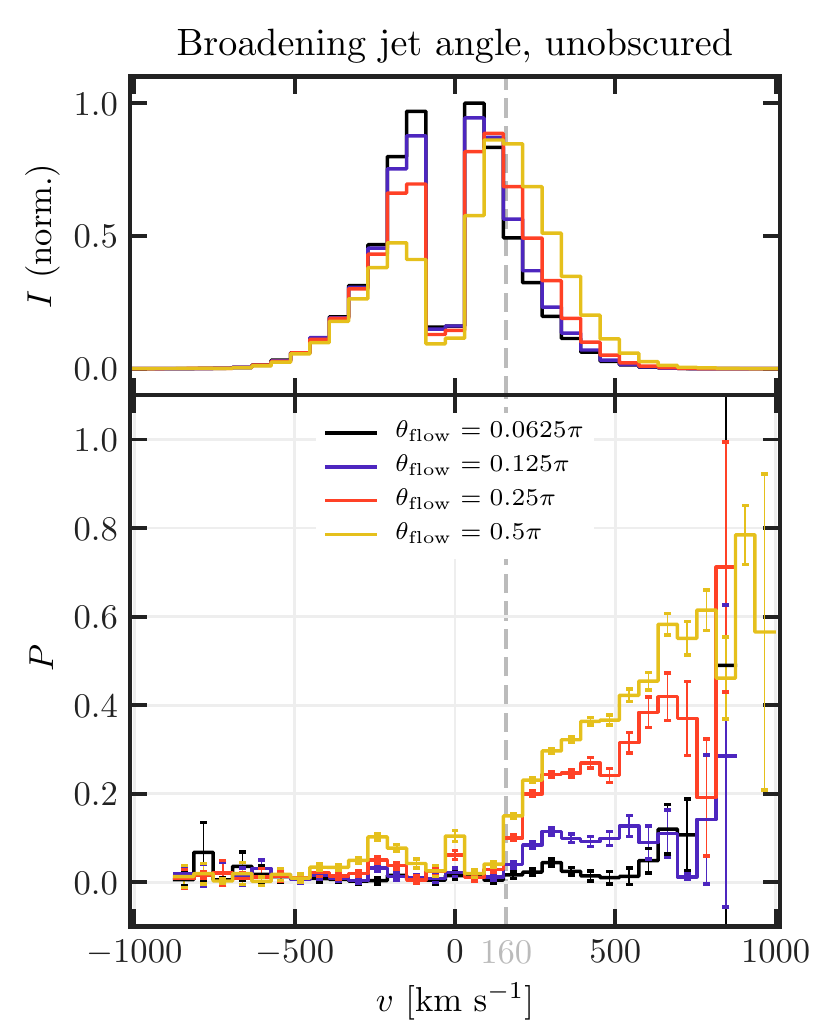}
    \,
    \includegraphics[width=\columnwidth]{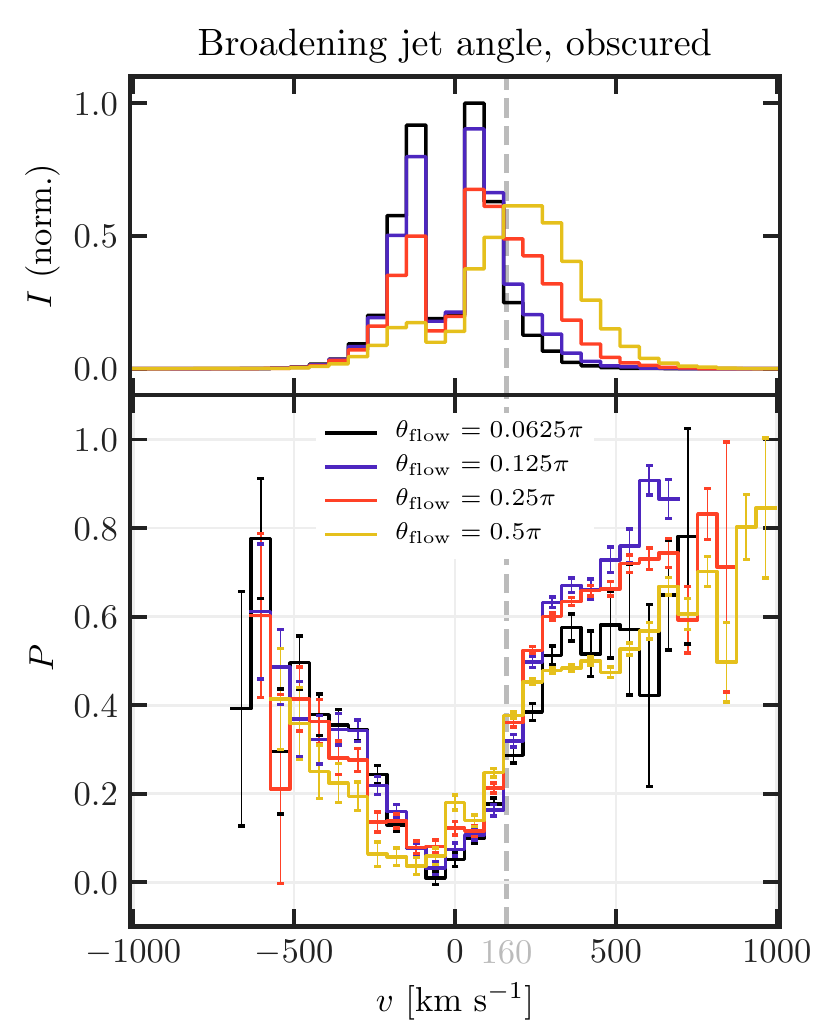}
    \caption{Intensity and polarization spectra for bipolar outflows with $v_{\rm flow} = 200$ km s$^{-1}$ out of a static HI non-elliptic sphere with radial column density $N_{\rm HI} = 10^{19}$ cm$^{-2}$. 
    The opening angle $\theta_{\rm flow}$ of the outflows out of the sphere is indicated by the line colors.
    The grey, vertical dashed lines indicate the velocity offset for which bluer photons would be removed by a partially neutral IGM.
    \textbf{\textit{Left}}: spectra for unobscured sources, ie.~including photons leaving both the central, static sphere and the moving medium in the bipolar cones. \textbf{\textit{Right:}} spectra for partially obscured sources, ie.~corresponding to blocking photons emitted within the horizontally dashed grey lines in Figure~\ref{fig:bipolar_polarimage}, resembling for example the removal of Ly$\alpha$ photons by a circumgalactic disk of dust. }
    \label{fig:bipolar_spectrum}
\end{figure*}

So far, our analysis has focused on spherically or cylindrically symmetric gas geometries. However, there is observational and theoretical evidence that outflows are bipolar \citep[e.g.][]{Blandford1974,Suchkov1994}. More recently, observations of LARS~05 \citep{Duval2016} nicely illustrate how Ly$\alpha$ photons scatter off a bipolar outflow that burst out of an edge-on disk galaxy. In this section, we focus on predicting spectra and polarization of scattered Ly$\alpha$ radiation emerging from simplified representations of bipolar outflows, with either an unobscured (\S~\ref{ssec:unobs}) or obscured (\S~\ref{ssec:obs}) central source.

We model the bipolar outflows as follows: it contains a spherical cloud with either $N_{\rm HI} = \{ 10^{17}, 10^{19}, 10^{21} \}$ cm$^{-2}$ and $T = 10^4$ K that resides in a fully ionized environment. The (unpolarized) Ly$\alpha$ source resides in the center of this cloud. We then introduce bipolar outflows in cones with total opening angles $\theta_{\rm flow} = \{ 1/16, 1/8, 1/4, 1/2 \} \pi$. 
Inside the cones, gas is radially outflowing with a constant velocity $v_{\rm flow} = 200$ km s$^{-1}$.
The HI number density in the cones is equal to that in the central sphere, and extend a factor of $4$ further than the edge of the sphere.

\subsubsection{Unobscured Central Source}\label{ssec:unobs}
Figure~\ref{fig:bipolar_polarimage} shows an illustrative example of the spatial distribution of intensity and polarization for an outflow with $\theta_{\rm flow} = \pi/8$ viewed edge-on\footnote{When 
viewing the outflows face on, i.e. straight into the cones and possibly also the central static cloud, one would observe a spherically symmetric source, and hence would any point source polarization signal be lost.}. 
We clearly see the biconical structure in both the intensity and the polarization images.
In the central, static sphere, the degree of polarization increases familiarly toward the limb, reaching $P \sim 30\%$ with the polarization angle oriented tangentially to the center.
The intensity decreases radially outward, both in the central cloud and in the cones.
This diffusion of photons gives rise to a decreasing surface brightness profile, see e.g.~Figure~4 of \cite{Dijkstra2006}.
In the outflows, the degree of polarization increases with $\Delta v$ reaching values of $P\sim 50\%$--$70\%$ (pixels with $P > 80\%$ exist, but the flux in these is negligible). The direction of polarization in the outflows are oriented perpendicular to the flow direction.

Figure~\ref{fig:bipolar_polarimage} also shows that the overall degree of polarization signal increases with the opening angle $\theta_{\rm flow}$ and is always aligned perpendicular to the outflow direction.
This is the same physical effect that we saw at play for the ellipsoids: the local radiation fields are stronger in the direction of the source, and are not isotropized, leaving a polarization vector tangential to the direction of the central source.

The dependence of integrated polarization on $\theta_{\rm flow}$ is summarized in Figure~\ref{fig:bipolar_matrix}, which shows clearly that $P$ increases with $\theta_{\rm flow}$ for all $N_{\rm HI}^{(c)}$. 
This increase reflects that a increasing $\theta_{\rm flow}$ causes a larger fraction of flux to emerge from the biconical outflows, thereby increasing the spatial asymmetry of the source.
Additionally, the biconical outflows appear as more polarized: fewer photons scatter here.
Those that do, propagate along the outflow direction, but have to scatter and escape at angles closer to 90$^{\circ}$ to reach the observer.
That way they gain large degrees of polarization that also are oriented perpendicular to the outflow axis.
We illustrate this in panel d) of Figure~\ref{fig:polarization_sketch}.

Figure~\ref{fig:bipolar_matrix} also shows that the overall polarization is maximal for $N_{\rm HI} = 10^{19}$ cm$^{-2}$. For $N_{\rm HI} = 10^{17}$ cm$^{-2}$, the central cloud is optically thin to most emitted Ly$\alpha$ photons, and there is little flux in the scattered component (and the flux that does scatter, scatters in the core which leads to a lower degree of polarization). The overall polarization for for $N_{\rm HI} = 10^{21}$ cm$^{-2}$ is lower because in this case, the scattering medium is optically thicker, which isotropizes the Ly$\alpha$ radiation field. In turn, this isoptropization reduces the overall polarization of the radiation that escapes.

The {\it upper left panel} of Figure~\ref{fig:bipolar_spectrum} shows the spectra of the models with $N_{\rm HI} = 10^{19}$ cm$^{-2}$ (i.e. with maximum polarization). The spectra are double-peaked, with the red peak becoming stronger relative to the blue peak with increasing  $\theta_{\rm flow}$. This increase reflects the increasing fraction of Ly$\alpha$ photons that scatter through the outflow. The {\it lower left panel} of Figure~\ref{fig:bipolar_spectrum} shows that the degree of polarization is negligible blueward of line center for all opening angles. Redward, the degree of linear polarization increases with $\Delta v$. 
This reflects that the blue peak consists of photons that escape from the static central cloud: polarization is canceled out from the symmetric central geometry (as for case \textit{a} in Figure~\ref{fig:polarization_sketch}).
This also holds to some extent for the red peak, but in addition, it has contributions from photons that have escaped into the biconical outflows. 
There, they scatter less and provide a higher overall local polarization signal.
The redward increase in the global polarization reflects the increasing spatial asymmetry with $\Delta v$.

\subsubsection{Obscured Central Source}\label{ssec:obs}
\begin{figure*}[t]
    \centering
    \includegraphics[width=\columnwidth]{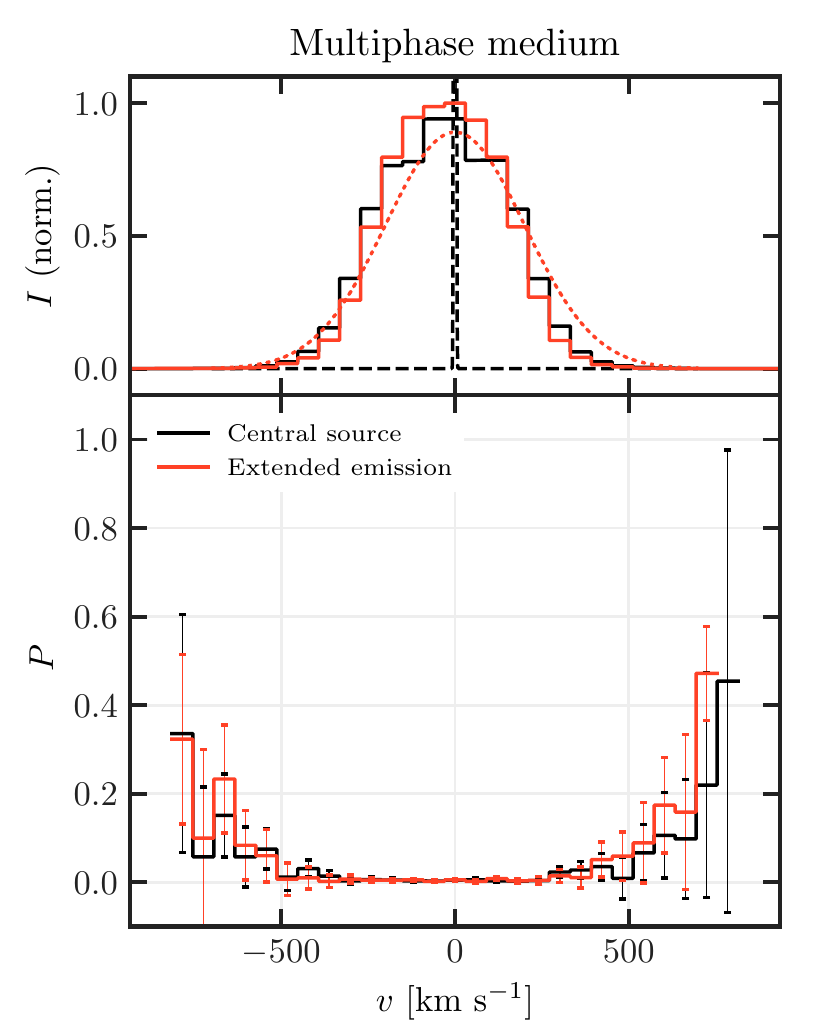}
    \,
    \includegraphics[width=\columnwidth]{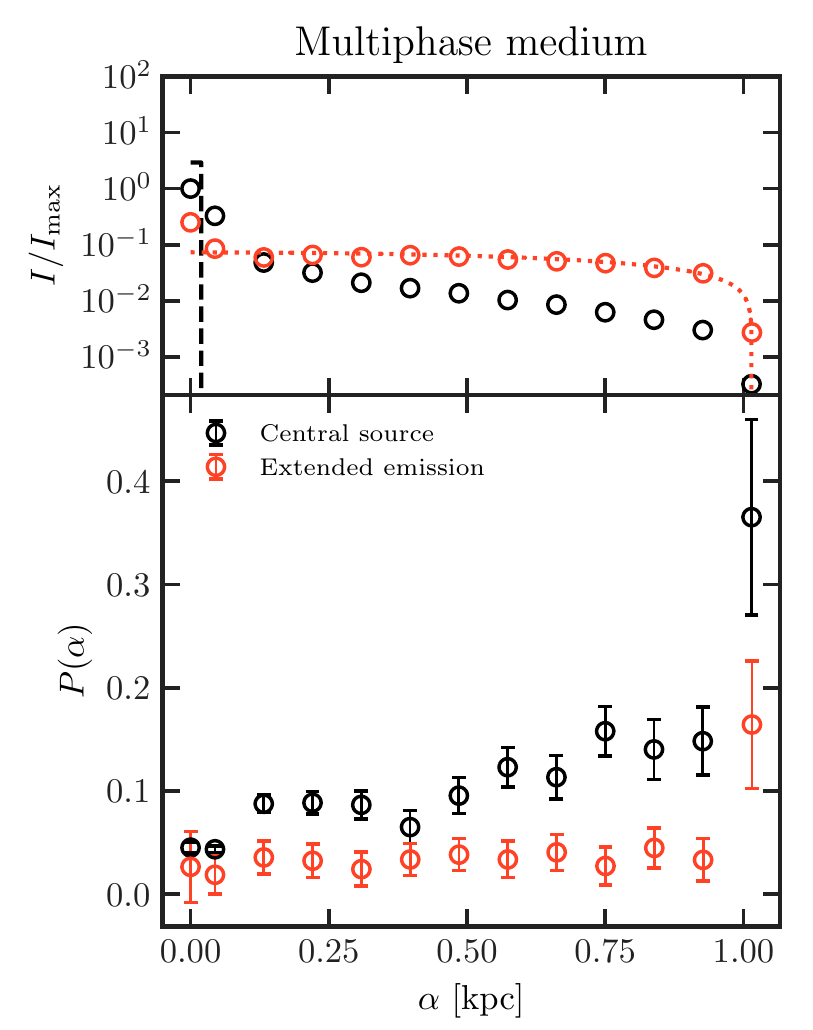}
    \caption{\textbf{\textit{Left:}} Spectra of intensity and degree of polarization as function of velocity offset/frequency from Ly$\alpha$ line center for systems filled
    with many small clumps of HI gas, sized to provide a covering fraction $f_{\rm c} \sim f_{\rm c,crit}$, in an otherwise ionized medium, representative of a multiphase scattering system. 
    We plot the emergent spectra for systems where the source of Ly$\alpha$ is central (black line), or where the source is extended throughout the medium (red line), ie., embedded in each cloud. 
    Note that we do not view the sources through a slit, but rather as point sources.
    We also plot the intrinsic intensity spectra, the dashed black line is for the central source, and the red dotted line is for the extended source.
        \textbf{\textit{Right:}} Surface brightness (normalized and rescaled to the maximum value obtained in the two models) and polarization profiles as function of impact parameter $\alpha$ for a central (black) and extend (red) source of Ly$\alpha$ emission in a clumpy, multiphase medium.
    We also plot the intrinsic surface brightness profiles, the dashed black line is for the central source, and the red dotted line is for the extended source.
    }
    \label{fig:clumpy_clouds}
\end{figure*}

We repeat the previous analysis (\S~\ref{ssec:unobs}), but now obscure the central static sphere (the obscured region is indicated with {\it grey dashed lines} in Figure~\ref{fig:bipolar_polarimage}). This represents a case in which the biconical outflows are separated by for example a dusty galactic disk as in LARS~05 \citep[see~][]{Duval2016} or in M82 \citep{Lynds1963,Gallagher1999}.

The {\it right panels} of Figure~\ref{fig:bipolar_spectrum} show the spectra \& polarization for the same models as in the {\it left panels}, but with the central region obscured.
Especially the red peaks of the spectra are widened for the largest opening angles $\theta_{\rm flow} = \{1/4, 1/2\}\pi$. 
This enhancement of the red peak is primarily a renormalization of the entire spectrum.
The obscuration removes a majority of the (blue and red) photons that arise from the central spherical cloud, leaving the surplus of red photons that escape from the cones.
For smaller opening angles $\theta_{\rm flow}$ however, less of the overall flux originates from the bipolar cones.
The surplus of red photons that was seen for the larger opening angles is present, but is however not sufficient to significantly alter the shape of the spectrum.
The spectral signature of the outskirts of the central sphere therefore dominate the spectrum.

The {\it lower right panel} shows that the polarization increases at effectively all frequencies. This simply reflects that obscuring the central source eliminates photons whose polarization vectors align with the cone axis.
The polarization in the blue wing is lower than in the red wing, as these are photons that escape from primarily the central sphere.
The additional boost in $P$ at large $\Delta v$ in the red wing is the signature of the photons that have scattered in the outflows.
As for the unobscured case, the local polarization is higher due to the fewer scatterings photons here undergo, and the global degree of polarization reflect the spatial asymmetry due to the cones.
The degree of polarization in the red wing reveals how the source transitions from nearly symmetrical without much contributions from the cones for $\theta_{\rm flow}=1/16$, to larger contributions from the cones with increasing $\theta_{\rm flow}$.
This comes at a price: the increased opening angles also allow for larger variations in the polarization vectors which in the cones are tangential to the source.

\subsection{Multiphase Medium}
\label{ssec:clouds}

All previous models represented gas in the ISM with a single density and temperature. In reality, interstellar (and circum-galactic) gas is known to be multi-phase. Ly$\alpha$ radiative transfer through multiphase media is a complex problem, which has also been represented by simplified models \citep[see e.g.~][]{Neufeld1991,Hansen2006,Dijkstra2012a,Laursen2013,Gronke2016}. These simplified models consist of neutral, spherical (possibly dusty) clumps, embedded within a hot, ionized and dust free medium \citep[based loosely on the early models by][]{McKee1977}.

It has been demonstrated that for such `clumpy' media, the key parameter that affects Ly$\alpha$ radiative transfer is the average 
number of clumps per sightline: the covering factor $f_{\mathrm{c}}$ \citep[see~][]{Hansen2006,Gronke2016}. 
\cite{Gronke2016,Gronke2017} showed that there exists a critical value for $f_{\mathrm{c}}$, $f_{\rm c, crit}$, above which clumpy media affect Ly$\alpha$ photons as if they consist of a single phase (i.e. homogeneous). The value of $f_{\rm c, crit}\sim$ a few - a few tens depending on the total HI column density and kinematics of the clumps (see Gronke et al. 2017). The polarization properties of Ly$\alpha$ radiation that scatters through `very clumpy' (i.e. $f_{\rm c} \gg f_{\rm c, crit}$) are therefore well captured by our previous models, in which the gas was homogeneously distributed. The polarization properties of Ly$\alpha$ radiation through models with $f_{\rm c} \ll f_{\rm c, crit}$ have been explored in \cite{Dijkstra2012a}, where it was associated with few (or no) scatterings and consequently high degrees of polarization. 

In this section, we focus on the `transition regime' which corresponds to $f_{\rm c} \sim f_{\rm c, crit}$, and contrast a central \Lya emitting source surrounded by a uniform distribution of randomly moving clumps (which can represent a central star forming galaxy surrounded by a clumpy circum-galactic medium) with a setup where the \Lya radiation emerges from the clumps (which can represent the same galaxy and circum galactic medium, but in which Ly$\alpha$ arises as fluorescent emission powered by ionizing radiation that leaked from the central galaxy, see \citealt{Mas-Ribas2016,Mas-Ribas2017}). While the numerical value of $f_{\mathrm{c, crit}}$ depends on the neutral hydrogen column density of the clumps $N_{\mathrm{HI, cl}}$ and their kinematics \citep{Gronke2017}, we stress that the characteristics described in this section apply generally for systems with $f_\mathrm{c} \sim f_{\mathrm{c,crit}}$.

In our models, we chose the clumps' column densities to be $N_{\rm HI,cl} = 10^{18}$ cm$^{-2}$ with a gas temperature of $T=10^4$ K \citep[motivated by the `shattering' theory of][]{McCourt2016}, and a random velocity with each component drawn from a Gaussian distribution with standard deviation $\sigma_{\rm cl} = 200$ km s$^{-1}$. This yields a critical covering factor of $f_{\mathrm{c,crit}}\approx 5$ \citep{Gronke2017} which we will adopt for $f_{\mathrm{c}}$. Furthermore, we choose the clumps' radii to be $r_{\mathrm{cl}} = 1\,$pc and fix the radius of the (spherical) system to $1\,$kpc. We note, however, that these parameters (given the others are fixed) do not influence the radiative transfer process \citep{Hansen2006}. We set the intrinsic spectrum to have $\sigma_i=12.85\,\mathrm{km}\,\mathrm{s}^{-1}$ (in the reference frame of the emitting gas) which corresponds to the thermal velocity of the gas.

The {\it upper left panel} of Figure~\ref{fig:clumpy_clouds} shows the spectra from a multiphase medium where \Lya is emitted either (a) by a central source or (b) extendedly, throughout the medium by sources residing in each clump. 
This is particularly visible in the intrinsic spectra, plotted with dashed and dotted lines. When having a central source, the photons are emitted close to the line center. With the extended source, the motion of the clumps must also be accounted for. Both spectra are broad, and single peaked, which is characteristic of media with $f_{\rm c} \sim f_{\rm c, crit}$ \citep[see Figure~3 of~][]{Gronke2017}.
Both models also predict a degree of spatially averaged polarization which is consistent with zero over most frequencies. 
This is a direct consequence of our symmetric scattering geometry---even if there exist patches that are tangentially polarized to the center, the overall geometry would cancel the global signal out, as illustrated in Figure~\ref{fig:polarization_sketch}.
The apparent rise in polarization toward the far wings occurs at frequencies with near zero intensity.

In the {\it right panel} of Figure~\ref{fig:clumpy_clouds}, we plot the normalized surface brightness profiles (which we after normalization render unitless by dividing with $I_{\rm max}$, the maximum surface brightness of the model with a central source) and polarization profiles of the sources as function of impact parameter $\alpha$ in kpc, in agreement with previous studies \citep[see~][]{Dijkstra2006}.
We also plot the intrinsic, unscattered surface brightness profiles with dotted/dashed lines.
In addition, the degree of polarization also differs. For the central source, $P$ rises to $\sim 10\%$ at $\sim 0.1$ kpc, before it eventually rises to $\sim 15\%$. For extended sources of Ly$\alpha$ emission, the polarization is consistent with $P<5\%$ out to the most distant impact parameters.
The degree of polarization is lower than that obtained for scattering off clumps at low $f_c < 1$ and greater $N_{\mathrm{HI, cl}}$ in \cite{Dijkstra2012a}, where photons that scattered only once in a clumpy outflow would produce a spatial polarization signal up to $P \approx 60$\% at large impact radii.
As we have a higher number of clumps along the line-of-sight (thereby a higher total optical depth) and a lower clump optical depth, the photons scatter several times per clump which reduces their polarization.
This explains the lower $P$ we obtain in the case of a central source as compared to \citet{Dijkstra2012a}.
However, we still obtain a similar increase in $P$ with impact radii as in other models with a central source since the radiation field is anisotropic, being stronger in the direction of the source, and the photons that escape at large radii must do so by scattering increasingly more at $90^\circ$.

This means, even though the exact degree of polarization depends on other parameters such as $N_{\mathrm{HI, cl}}$ and the clump placement, the central source shows a rising $P(r)$ signal -- while the ``flourescent'' clumps, i.e., the extended source does not. This is a clear observational signal for the distinction of different \Lya powering mechanism, and we will explore this further in future work.

\section{Discussion}
\label{sec:discussion}

In this section, we discuss the origins of polarization, from quantum mechanical (\S~\ref{ssec:disc_QM}) to astrophysical scales (\S~\ref{ssec:disc_intuitive}). We discuss how \Lya polarization can break degeneracies between models for spectra and/or surface brightness profiles, when used \textit{in concert} with these other observables (\S~\ref{ssec:disc_degeneracies}).

\subsection{Polarization: the Quantum Mechanical Origins}
\label{ssec:disc_QM}

\begin{deluxetable}{l  c c | c c }
\tablecaption{Polarization through single scatterings.
\label{tab:QM_polarization}
}
\tablecolumns{5}
\tablehead{
\colhead{} \vspace{-0.4cm} &
\multicolumn{2}{c}{\textbf{Core}} & 
\multicolumn{2}{c}{\textbf{Wing}} \\
\colhead{\bfseries Init.~polarization} \vspace{-0.4cm} &
& & &  \\
\colhead{} &
$90^\circ$   & 
$0$ /$180^\circ$  & 
$90^\circ$  & 
$0$ / $180{}^\circ$
\\ \vspace{-0.4cm}
}
\startdata
\sidehead{} \vspace{-1.3cm} \\
\textit{Unpolarized \tablenotemark{a}} \vspace{-0.9cm}   & 43\%         & 0\%     & 100\%       &                     \\
    \\ \vspace{-0.25cm} &&&& \textit{unchanged} \\
\textit{Polarized \tablenotemark{b}}                     & 43\%         & 60\%    & 100\%       & 
    \\[-0.75cm]
    \sidehead{}
\enddata
\tablecomments{\textit{Core} scattering in this regard includes only the anisotropic K transition, and not the depolarizing H transition.}
\tablenotetext{a}{An \textit{initially} unpolarized photon has $P=0\%$.}
\tablenotetext{b}{An initially polarized photon conversely has $P=100\%$.}
\end{deluxetable}

For classical electron-scattering - which applies to Ly$\alpha$ wing scattering - unpolarized radiation that scatters at right angles become maximally polarized. In addition, the polarization properties of a single photon impose restrictions on the scattering angles: the photon cannot scatter in the direction in which it is fully polarized.
For core scattering however, the `shape' of quantmum-mechanical wavefunctions plays a role. Unpolarized radiation still obtains the highest degree of polarization when scattering at right angles, but only up to $P=43\%$ via K scatterings (i.e. through the 2P$_{3/2}$ state). For H scatterings (i.e. through the 2P$_{1/2}$ state) polarization is only destroyed as the wavefunction of the 2P$_{1/2}$ is spherically symmetric. For fully polarized radiation, i.e. with $P=100\%$,  a photon that is H scattered will only obtain $P=43\%$ for scattering at right angles, but for forward or backward scatterings, the degree of polarization can increase to $P=60\%$. This effect in known as \textit{depolarization}: a photon can only obtain $P=100\%$ through wing scatterings, and if it is scattered through the core, it will only retain, at best, $P=60\%$. Depolarization is not possible through wing scatterings: a partially polarized photon that is forward or backward scattered will retain its polarization, or have it boosted when scattering at inclined angles.

We obtain these polarization magnitudes by using the desired scattering angles for the density matrices of the various transitions. These values agree perfectly with the results for Rayleigh or core scatterings as described, e.g., in \cite{Chandrasekhar1960} or \cite{Dijkstra2008}.
We summarize this discussion in Table~\ref{tab:QM_polarization}
which provides an overview of the polarization obtained through single scattering.

\subsection{Polarization: the Astrophysical Origins}
\label{ssec:disc_intuitive}
There is a difference between the probability of measuring the polarization state of an individual photon---which we just have shown can grow to prefer highly polarized photons after multiple scatterings---and observationally detecting polarization. 
The observable Stokes parameters describes the polarization properties of an \textit{ensemble} of photons. While individual photons may obtain high levels of linear polarization through scattering, the ensemble averaged polarization can still be zero -- if their polarization angles are not well aligned.

There are two main mechanisms behind the alignment of the polarization vectors, and thus, of generation of observable polarization: \textit{natural asymmetries in the scattering geometry}, or \textit{introduced asymmetries} from e.g.~finite slit widths in spectropolarimeters, foregrounds or instrumental artefacts. We discuss these next.\\

\textbf{I. Natural asymmetries}: On scales where the properties of the scattering medium appears constant (local scales), any process that induces some preference in scattering direction, also introduces a preferential polarization direction. The polarization can not be oriented in the direction the photon had before scattering and it must also be perpendicular to the post-scattering propagation direction. Such a process can be an alignment of the atoms in the medium either from an external magnetic field or pumped by scatterings \citep[see e.g.~][]{Zhang2017}, or as in our cases, an anisotropic radiation field, as also realized by \cite{Dijkstra2008}. A small, local patch of the scattering medium is in general unevenly illuminated, with strongest illumination in the direction of the source. This is the origin of the tangential polarization patterns one would obtain from scatterings in the expanding IGM \citep{Rybicki1999}, in spherical shells \citep{Dijkstra2008}, or in any of the symmetric regions of our scattering models including the spherical ellipsoid ($\varepsilon=1/1$) in \S~\ref{ssec:ellipsoid}, the centrally illuminated clumpy medium in \S~\ref{ssec:clouds} with radial polarization profile given in Figure~\ref{fig:bipolar_spectrum} or the IGM (radial profile in Figure~\ref{fig:polardegree_tests}). Observationally, such a polarization pattern of concentric circles was observed in LAB1 by \cite{Hayes2011}, illustrated in \cite{Bower2011}. 
At larger impact radii, these models possess a larger degree of polarization.
This is due to the larger fraction of photons escaping at right angles with increasing impact parameter.
As was shown in the previous section (also see Table~\ref{tab:QM_polarization}), this is accompanied by higher degrees of polarization.
However, when the systems are symmetrical, their net polarization will cancel out, as illustrated in the lower row of panel a) in Figure~\ref{fig:polarization_sketch}.

This brings us to our next important realization: on a macroscopic, global level, scattering through a geometrically \textit{asymmetric} system can result in polarization, as also found by \cite{Angel1969} and \cite{Lee1998}
This polarization is detectable even \textit{without} spatially resolving the system.
We have shown this by introducing ellipticity, biconical outflows and central disk-like obscuration.
In panels b), c) and d) of Figure~\ref{fig:polarization_sketch}, we display examples of such asymmetric systems and their overall polarization signatures.
A spatially-averaged polarization signal requires, in the first place, that the local radiation field is polarized.
The global polarization orientation is then tangential to the source and reflects the location of the asymmetric regions that provide the surplus of polarized photons. For example, in the bipolar outflow, they emerge from the cones, and the polarization direction is consequently perpendicular to the outflow direction (see \S~\ref{ssec:outflows}). This is similar to the observations of the Egg nebula by \cite{You2017}.
When the local radiation fields were isotropized by a high number of core scatterings, we found, as \cite{Lee1994a} and \cite{Dijkstra2008} that this reduces the emergent polarization.
An example of this is an outflowing, oblate system, in which an increased column density could reduce the polarization as well as flip the polarization vector (see~Figure~\ref{fig:matrix_polarization_expanding_ellipsoids}). 
The same effect occurs in the multiphase systems studied (\S~\ref{ssec:clouds}), i.e., several core scatterings lead to a decrease in polarization -- as also identified by \cite{Dijkstra2008} in the context of intergalactic propagation.

The above asymmetries are purely \textit{geometrical}. 
It is also possible to introduce asymmetries in \textit{velocity space}
since a velocity field can lower the (frequency dependent) optical depth of a systemt -- with similar effects as described above.
Examples are the ellipsoids presented in \S~\ref{ssec:ellipsoid} that became stronger polarized in the presence of global outflows, and the outflows in the biconical structures in \S~\ref{ssec:outflows} which allowed for scattering in them.\\

\textbf{II. Introduced asymmetries}: These occur when observing a patch of a larger geometry (intentionally or unintentionally), that is, by effectively masking out regions which would alter the observable.
This is illustrated by the sketch in Figure~\ref{fig:polarization_sketch}a. With a slit or aperture covering the entire system, one would detect no polarization as the symmetric polarization vectors cancel out (illustrated in {\it the lower panel}). However, if we observed part of the system through a narrow slit, then polarization contributions outside the slit are removed, which breaks the symmetry, and yields a surplus of polarization perpendicular to the slit alignment direction. This would result in a global (but possibly misleading) polarization signal.
Of course, these issues are less important in imaging polarimetry when one can obtain Stokes parameters in a per-pixel basis (as in \citealt{Hayes2011}, \citealt{Prescott2011} and \citealt{You2017}).

\subsection{Polarization: Breaking Degeneracies}
\label{ssec:disc_degeneracies}
\begin{figure}[t]
    \centering
    \includegraphics[width=\columnwidth]{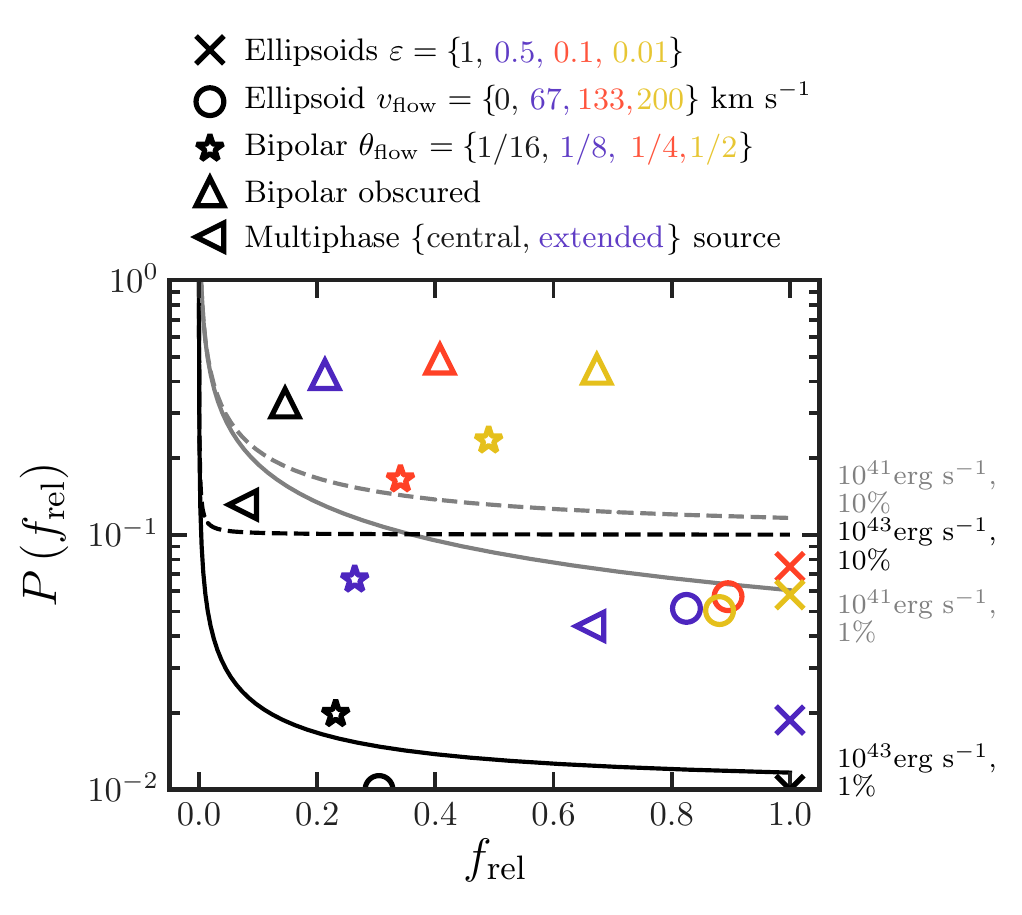}
    \caption{Detectability of models studied in this work.
    We show the degree of polarization at those frequencies/radii where $P$ differs among them (with $f_{\rm rel}$ giving the fraction of the total flux for these, see \S~\ref{ssec:disc_degeneracies}).
    The solid and dashed lines show the $1\sigma$ detection limit for a $z \sim 3$ source observed for 1 hour with a VLT-like telescope. We show the impact of a 1\% (10\%) systemic error, and a different luminosity, both shown as labels on the right-hand side of the plot.}
    \label{fig:observability}
\end{figure}

We have shown in the results that polarization signals themselves can be degenerate, i.e., several setups can produce similar polarization signatures. A prime example of this is the global polarization signals obtained from non-static asymmetric scattering geometries (our ellipsoids, see Figure~\ref{fig:matrix_polarization_expanding_ellipsoids}).
Here, the polarization angle \textit{flips} from being perpendicular to the major axes of the system, to being parallel to it, when the column density is increased. Also, when further thinning the medium by introducing outflows, the degree of polarization \textit{decreases} at higher column densities, compared to an \textit{increase} with lower column densities.

The origins of this flip is the change of apparent geometry in different column densities (the transition from column \textit{b} to \textit{c} in Figure~\ref{fig:polarization_sketch}), by lowering the optical depth, one also transitions from escape and scattering mainly along the minor axis to scattering and escaping in the full system. In the latter case, a surplus of photons escape from the extended lobes, being polarized tangentially to the source, and perpendicular to the major axes.

Without knowledge of the apparent geometry of a system (which is generally the case), we cannot solely use the degree and direction of polarization to constrain the major axis of the system (compare the case  $N_\HI^{(c)} = 10^{19}$ cm$^{-2}$ and $v_{\rm exp} = 0$ km s$^{-1}$ to that of rotated system with $N_\HI^{(c)} = 10^{21}$ cm$^{-2}$ and $v_{\rm exp} = 200$ km s$^{-1}$ in Figure~\ref{fig:matrix_polarization_expanding_ellipsoids}). Similarly, we cannot differentiate between a strong bipolar flow-like geometry in which the polarization arises due to scattering in lobes, and a more compact - slightly asymmetric - system with obscuration of the core (our \S~\ref{ssec:ellipsoid} compared to \S~\ref{ssec:outflows}): In both cases, we would have a polarization signal aligned with the major axis of the system.

However, similar degeneracies also exist when using \textit{other observables}: The spectrum is most sensitive to the properties of the scattering medium along the path of least resistance. The ellipsoids explored in \S~\ref{ssec:ellipsoid} and \S~\ref{ssec:exp_ellipsoid} show examples of this: the spectra do not change when the system changes from being viewed face-on to edge-on. In addition, Ly$\alpha$ spectra do not necessarily reveal intrinsic dynamics of the scattering gas. Scattering through outflows gives rise to asymmetric spectra, often with a negligible blue peak as in Figure~\ref{fig:spectra_expanding} and \ref{fig:bipolar_spectrum}. In the same figures, and in Figure~\ref{fig:spectra_ellipsoids}, we have plotted vertical dashed gray lines, which mark the range of frequencies which could suppressed by scattering in the intergalactic medium \citep{Dijkstra2007,Laursen2011}, and leave a spectral signature virtually identical to that generated by scattering through an optically thick, outflowing medium. This illustrates that degeneracies can exist when using spectra alone.

A joint analysis of \Lya observables can break the mentioned degeneracies: the spectrum can constrain the column density of the optically thinner minor axis. We can then use polarization to constrain the orientation of the system.
This allows, for instance, to differentiate between the polarization signals from static, asymmetric systems to those of dynamic, possibly geometrically symmetrical, systems. In static gas geometries, scattering gives rise to symmetric spectra, whereas scattering through dynamic gas geometries generally gives rise to asymmetries in the spectra. Another example relates to which processes removes flux blueward of a galaxies systemic velocity in \Lya spectra: IGM or outflows. The IGM can transform an intrinsically double peaked profile emerging from a static medium into a spectrum with a dominant red peak, and can therefore mimick the effect of scattering through a galactic outflow. However, with only outflows, the polarization increases with offset from the line center (as seen in Figures~\ref{fig:spectra_expanding} and \ref{fig:bipolar_spectrum}) while it does not in static systems (see Figure~\ref{fig:spectra_ellipsoids}).
A static system with the IGM processing away the blue peak would hence give rise to different polarization signature. In the case of \textit{both} IGM processing and outflows, the \textit{degree} of polarization can be attempted reconciled with the observed spectral shape.

Finally, in Figure~\ref{fig:observability} we quantify the ability of present-day telescopes to differentiate between polarization signatures of all our explored models.
In order to do this, we define the fraction of the total flux $f_{\rm rel}$ where the polarization signal differs. This fraction can be defined spatially (e.g., only photons arriving in the outer regions for the multiphase media in \S~\ref{ssec:clouds}) or in frequency space. This corresponds to an optimally designed experiment where, for instance, the slit position has been chosen so that only photons with a positive net polarization are recorded. 
Specifically, we then the show the degree of polarization for this fraction of photons $P(f_{\rm rel})$ versus $f_{\rm rel}$ in Figure~\ref{fig:observability}.

For the outflowing ellipsoids and bipolar outflows (unobscured and obscured), $f_{\rm rel}$ is obtained for $v > 160$ km s$^{-1}$ (also indicated by vertical dashed gray lines in the spectra in Figures~\ref{fig:spectra_expanding} and \ref{fig:bipolar_spectrum}), where the polarization differ the most. For the static ellipsoids, the polarization differ similarly across the spectrum, and all frequencies are included, hence $f_{\rm rel} = 1$ for those.
We used photons arriving from $r>0.1 R_{\rm max}$ for the multiphase media with either an central source of \Lya or sources extending throughout the medium as the frequency-dependent polarization for both those models were near zero.
With the solid lines, we show the sensitivity\footnote{We restrict ourselves to a shot-noise/systemic-limited approximation $\sigma_P = \sqrt{1/\left((N_{\rm HWP}/2) {\rm SNR} \right)^2 + \sigma_{\rm syst}^2}$ where $N_{\rm HWP}=4$ is the number of half-wave plate rotations, ${\rm SNR} = \sqrt{N_{\rm phot}}$ is the signal-to-noise ratio given in the case of shot-noise only from the number of photons $N_{\rm phot}$ arriving at the sensor, and $\sigma_{\rm syst}$ is a systemic error. Based on \cite{Patat2006}.} of FORS2 at VLT. 
We see that it would be able to differentiate between most models if these were $L=10^{43}$ erg s$^{-1}$, or even $L=10^{41}$ erg s$^{-1}$, \Lya emitters at $z=3$ that were observed for one hour.
Separating models becomes harder if one assumes that the systemic error of the instrument is as high as 10\%, except for those models where $P({\rm frel})>20\%$.
Detecting, differentiating $P$ and breaking the otherwise degenerate models we have explored is thus viable already today.
 
\begin{deluxetable*}{l  c c | c c | c c | c c }
\tablecaption{Intensity and polarization properties of \Lya systems explored in this paper.
\label{tab:summary}
}
\tablecolumns{9}
\tabletypesize{\footnotesize}
\tablehead{
\colhead{} &
\multicolumn{2}{c}{\S~\ref{ssec:ellipsoid} \textbf{Ellipsoid}\tablenotemark{a}} & 
\multicolumn{2}{c}{\S~\ref{ssec:exp_ellipsoid} \textbf{Ellipsoidal outflow}\tablenotemark{a}} &
\multicolumn{2}{c}{\S~\ref{ssec:outflows} \textbf{Bipolar outflow}\tablenotemark{a}} &
\multicolumn{2}{c}{\S~\ref{ssec:clouds} \textbf{Multiphase medium}\tablenotemark{b}} \\
\colhead{} &
\colhead{Spherical} &
\colhead{Ellipsoidal} &
\colhead{Low $v_{\rm flow}$} &
\colhead{High $v_{\rm flow}$} &
\colhead{Small $\theta_{\rm flow}$} &
\colhead{Large $\theta_{\rm flow}$} &
\colhead{Central source} &
\colhead{Extended source}
\\ \vspace{-0.4cm}
}
\startdata
\sidehead{} \vspace{-1.1cm} \\
\textbf{$I(v)$}   &   \multicolumn{2}{c|}{Symmetric double-peaked} & \multicolumn{2}{c|}{Redshifted single-peaked} & \multicolumn{2}{c|}{Redshifted doubled-peaked} & \multicolumn{2}{c}{Broad, single peak}    \\
\textbf{$\langle P \rangle$}\tablenotemark{1}   &   Zero\tablenotemark{2} & $\sim 5\%$ & \multicolumn{2}{c|}{Depends on $N_{\rm HI}$} & 1\% & 10\% & Zero\tablenotemark{2} & $<5\%$ also locally   \\
\textbf{$P(v)$}   & Flat, nil & Flat, non-zero & \multicolumn{2}{c|}{Rises\tablenotemark{3} to $\sim$ $30\%$} & Flat, low & Rises up to $60\%$ & \multicolumn{2}{c}{Flat, zero} \\
\textbf{$P(r)$} & \multicolumn{2}{c|}{Rising} & \multicolumn{2}{c|}{Rising} & \multicolumn{2}{c|}{Rising} & Rising & Flat, zero \\
Figures        & \multicolumn{2}{c|}{\ref{fig:spectra_ellipsoids}, \ref{fig:matrix_polarization_ellipsoids}} & \multicolumn{2}{c|}{\ref{fig:spectra_expanding}, \ref{fig:matrix_polarization_expanding_ellipsoids}} & \multicolumn{2}{c|}{\ref{fig:bipolar_polarimage}, \ref{fig:bipolar_matrix}, \ref{fig:bipolar_spectrum}} & \multicolumn{2}{c}{\ref{fig:clumpy_clouds}}
    \\[-0.65cm]
    \sidehead{}
    \enddata
\tablecomments{The exact numerical values are model dependent.
}
\tablenotetext{a}{Viewed edge-on.}
\tablenotetext{b}{Our clumpy clouds have a covering fraction close to the critical value, $f_{\rm c} \sim f_{\rm c, crit}$.  Other $f_{\rm c}$ produce different spectra.}
\tablenotetext{1}{The global, frequency- and spatially-integrated polarization values are given as the extremes obtained for the models.}
\tablenotetext{2}{The polarization is \textit{locally} non-zero, however. Global symmetries cancel it out as illustrated in panel a) of Figure~\ref{fig:polarization_sketch}.}
\tablenotetext{3}{These increases in $P$ with $v$ depends on the column density of the system and are here given for $N_{\rm HI} = 10^{19}$ cm$^{-2}$ along the minor axes.}
\end{deluxetable*}

\section{Summary and Conclusions}
\label{sec:conclusion}
A major challenge in extragalactic astrophysics is to decode and reveal the properties of systems with only a limited set of observables.
Interpreting observations of \Lya requires us to understand the transport of this radiation. Spectra and/or surface brightness profiles provide constraints on this scattering process, though sometimes not uniquely so.
The polarization properties of Ly$\alpha$ provide additional constraints on the scattering process, but this has been explored much less in the literature, even though it has been demonstrated that extragalactic sources of Ly$\alpha$ source can reach significant degrees of polarization (both theoretically and observationally).

This motivated us to implement polarization into the \Lya radiative transfer code \texttt{tlac} of \cite{Gronke2014}, providing us with the ability to do a joint analysis of emergent observables.
To this aim, we used the \textit{density matrix formalism} of \cite{Lee1994}. 
Through a $2 \times 2$ matrix, it properly describes the probability of measuring a \Lya photon in either of its two helical spin states, as well as the linear superposition of these.
The elements of this are modified through (\Lya) \textit{core scatterings} near the line center and \textit{wing scatterings} in the damping wing.
As photons escape from an arbitrary three-dimensional \HI scattering medium that contain a single or a distribution of \Lya sources, we convert the density matrix coefficients of each individual photon to observable Stokes parameters.
This approach allowed us to treat polarization both on a quantum mechanical level as well on a statistical, observable level, setting this work apart from earlier works where only the latter could be achieved (as in~\citealt{Rybicki1999}; \citealt{Dijkstra2008} and \citealt{Trebitsch2016}).

We have explored scattering through a suite of simplied geometries with simplistic dynamics, such as static and expanding ellipsoids, biconical outflows, and multiphase (clumpy) outflows. We summarize some of their observable properties in Table~\ref{tab:summary}.
These idealized models help in understanding the physical origins of the polarization signal, and correspond to simplified setups for which other \Lya observables have been studied previously in the literature. 
We have shown how the global signal from an unresolved source, either its degree and angle of polarization, or its polarization spectrum, depends on its scattering symmetry.
A symmetric system would appear to have zero polarization, just as one that either is sufficiently optically thick to isotropize the emergent radiation, or that emits \Lya extensively \textit{throughout} itself (as in cooling systems, or in the case of recombinations/fluoresence) would have.

Introducing asymmetries in the scattering geometry, from the smallest to the largest scales, we showed how polarization is generated.
The polarization is a measure of any surplus, or lack of, scattered photons at locations for which the scattering geometry is not fully symmetric.
We explored ellipsoids and bipolar outflows as examples of this.

Alone, the polarization signal cannot -- just as other observables -- be used alone to describe the physical state of a source and its environment.
We have shown that this is only possible when it is used in conjunction with other observables.
As an example, we have shown that we obtain tangential polarization patterns around central sources. 
In asymmetric geometries, the polarization direction may be used to reveal the alignment of the system. 
But this only works if the intensity spectrum of the system is known, as the polarization direction is degenerate between several geometries and dynamics.

Likewise, other observables should be used with caution.
Knowing an intensity spectrum or luminosity of a source, one may misinterpret these as being intrinsic to the source, although in an asymmetric system with anisotropic \Lya escape, it is not necessarily so.
The emergent intensity only reveals the properties of the medium along the path of least resistance.
With polarization arising due to asymmetries, a global polarized \Lya signal of an unresolved source would be a strong indicator of possible anisotropies in the \Lya escape.
In systems with anisotropic \Lya escape, it is easy to misinterpret this as a low \Lya (and in some cases, LyC) escape fraction.
Moreover, IGM absorption manifests itself through an attenuation of the blue part of the intensity spectra of sources, but this could also be falsely mistaken to be caused by outflows, which also imprint this spectral signature.
We have shown that it is possible to break this degeneracy with polarization measurements.

Currently, none of the next generation extremely large telescopes plan to include polarimeters intended for extragalactic use \citep[see discussion in][]{Hayes2011a}.
However, new, dedicated observations are undertaken \citep[see e.g.][]{Beck2016,You2017}, promising a bright future. 
We have also shown that present-day telescopes would be able to differentiate between the polarization signals of most of the models we have explored.
We are currently exploring a realistic, multiphase medium in an upcoming paper, comparing it to recently obtained observations.
Continued work is needed, both theoretically \citep[see e.g.~][]{Chang2017}, numerically and observationally, as we have shown in this paper that polarized \Lya can be a powerful, degeneracy-breaking probe into an otherwise secretive Universe.

\acknowledgments
We all thank B.~Ciardi for helpful comments.
We thank the referee for the highly constructive feedback.
MBE is grateful to H.-W.~Lee, S.-J.~Chang, C.~Scarlata, C.~You and Ll.-M.~Ribas along with other members of the observational and theoretical Ly$\alpha$ community (you know who you are) for rewarding discussions.
MBE thanks the Institute of Theoretical Astrophysics at UiO and the Astronomy and Astrophysics Department at UCSC for their kind hospitality.
MD thanks the physics department at UCSB for their kind hospitality.
MH acknowledges the support of the Swedish Research Council, Vetenskapsr{\aa}det and the Swedish National Space Board (SNSB), and is Fellow of the Knut and Alice Wallenberg Foundation.

\software{\texttt{tlac} \citep{Gronke2014}, \texttt{numpy} \citep{VanderWalt2011}, \texttt{Cython} \citep{Behnel2011}, \texttt{Matplotlib} \citep{Hunter2007}.}

\bibliographystyle{yahapj}
\bibliography{references}

\appendix
\section{Density matrices for \Lya transitions}
\label{app:density_matrices}
Here, we will present the elements of the $2 \times 2$ density matrices describing the probability of measuring the polarization state of a photon following scattering through the three possible \Lya transitions, following \cite{Ahn2002} or \cite{AHN2015POLARIZATIONHYDROGEN}. 

In the following, we use the left-handed photon basis $\left\{ \boldsymbol{\varepsilon}_1, \boldsymbol{\varepsilon}_2, \boldsymbol{\varepsilon}_3\right\}$ where
\begin{align}
    \boldsymbol{\varepsilon}_1 &= \left( -\sin \phi, \cos \phi, 0 \right) \label{eq:epsilon_1} \\
    \boldsymbol{\varepsilon}_2 &= \left( \cos \theta \cos \phi, \cos \theta \sin \phi, -\sin\theta \right) \label{eq:epsilon_2} \\
    \boldsymbol{\varepsilon}_3 &= \left( \sin \theta \cos \phi, \sin \theta \sin \phi, \cos \theta \right) \label{eq:epsilon_3}
\end{align}
with $\phi$ and $\theta$ as the polar angles.

The state of a \Lya photon after scattering is determined by eight unique photon parameters: (\textit{i}) the photon frequency determining whether the transition occurs in the resonance core or the damping wing and hence the transition type; (\textit{ii}--\textit{iii}) the two polar angles $\theta$ and $\phi$ uniquely specifying the photon orientation prior to scattering; (\textit{iv}--\textit{v}) the polar angles $\theta'$ and $\phi'$ following a scattering; and (\textit{vi}--\textit{viii}) the density matrix of the unscattered photon $\rho_{\rm phot}$ with unique elements $\rho_{\rm phot,11}, \rho_{\rm phot, 12}=\rho_{\rm phot, 21}^*=\rho_{\rm phot, 21}, \rho_{\rm phot, 22}$.
A difference in angles is given as $\Delta \phi = \phi' - \phi$.
The state of the photon after a scattering can then be represented by a new density matrix $\rho_{\rm phot}'$.

For H-type resonance core scatterings, through the $2P_{1/2}$ state of \HI, in which the wave-function of the hydrogen atom has no angular dependence, the photon is depolarized following a scattering event and the resulting density matrix elements are
\begin{equation}
    \rho_{11}' = \rho_{22}' = 1/2; \quad \rho_{12}' = 0.
    \label{eq:density_matrix_H}
\end{equation}

For K-type resonance core scatterings, through the $2P_{3/2}$ state, there is a strong angular dependence and the polarization state of the incoming photon,
\begin{align}
    \rho_{11}' &= \left( 5+3 \cos 2\Delta\phi \right) \rho_{11} 
                - 6 \cos\theta \sin 2\Delta\phi \rho_{12}
                + \left[ \left( 5-3 \cos 2\Delta \phi \right) \cos^2 \theta 
                        + 2 \sin^2 \theta \right] \rho_{22} 
    \label{eq:K_11}
\end{align}
\begin{align}
    \rho_{12}' &= 3 \sin 2 \Delta \phi \cos\theta' \rho_{11} \notag\\
    &+ 6 \left[ \cos\theta \cos\theta' \cos 2\Delta\phi 
                         + \sin\theta \sin\theta' \cos\Delta\phi \right] \rho_{12} \notag\\
    &+ 3 \cos \theta \left( -2 \sin\theta\sin\theta' \sin\Delta\phi 
                                - \cos\theta\cos\theta' \sin2\Delta\phi \right) \rho_{22}
    \label{eq:K_12}
\end{align}
\begin{align}
    \rho_{22}' &= \left[\left( 5-3 \cos2\Delta\phi \right) \cos^2\theta'
                        + 2 \sin^2\theta' \right]\rho_{11} \notag\\
               &+ \left[ \left( 5+3 \cos2\Delta\phi \right) \cos^2\theta \cos^2\theta'
                        + 2\cos^2\theta \sin^2\theta' 
                        + 12 \cos\Delta\phi \cos\theta' \cos\theta \sin\theta \sin\theta'
                        + 2 \cos^2\theta' \sin^2\theta + 8\sin^2\theta \sin^2\theta'\right] \rho_{22} \notag \\
               &+ \left( 6 \sin 2\Delta\phi \cos\theta \cos^2\theta'
                        +2 \sin\Delta\phi \cos\theta' \sin\theta \sin\theta' \right)\rho_{12}.
    \label{eq:K_22}
\end{align}

Whereas for wing scatterings, in which the \cite{Stenflo1980}-effect effectively make the transition behave as scattering off a free electron, the resulting density matrix elements are:
\begin{align}
    \rho_{11}' &= \cos^2\phi \rho_{11} - \cos\theta \sin 2\Delta\phi \rho_{12} + \cos^2\theta \sin^2\Delta\phi \rho_{22}
    \label{eq:wing_11}
\end{align}
\begin{align}
    \rho_{12}' &= 1/2 \cos \theta' 2\Delta\phi \rho_{11}  \notag\\
               &+ \left[ \cos\theta \cos\theta' \cos 2\Delta\phi + \sin\theta\sin\theta' \cos\Delta\phi \right] \rho_{12} \notag\\
               &- \cos \theta \left( \sin\theta \sin\theta' \sin\Delta\phi + 1/2 \cos\theta\cos\theta' \sin 2\Delta\phi \right)\rho_{22}
    \label{eq:wing_12}
\end{align}
\begin{align}
    \rho_{22}' &= \cos^2\theta' \sin^2\Delta\phi \rho_{11} \notag\\
    &+ \cos\theta' \left[ 2\sin\theta\sin\theta' \sin\Delta\phi
    + \cos\theta\cos\theta' \sin 2\Delta\phi \right] \rho_{12} \notag\\
    &+ \left( \cos\theta \cos\theta' \cos\Delta\phi + \sin\theta\sin\theta' \right)^2 \rho_{22}.
    \label{eq:wing_22}
\end{align}

\section{Translating to an observer-specific coordinate system}
\label{app:translating_to_observer_basis}
The polarization vectors $\boldsymbol{\varepsilon}_1$ and $\boldsymbol{\varepsilon}_2$ chosen by \cite{AHN2015POLARIZATIONHYDROGEN} were specific for a semi-infinite planar slab in the $x$, $y$-directions. The density matrix elements are composed of the (possibly complex) coefficients $c_1$ and $c_2$ which were estimated for the chosen polarization vectors $\boldsymbol{\varepsilon}_1$ and $\boldsymbol{\varepsilon}_2$, which also are given in Eqs.~\ref{eq:epsilon_1} and \ref{eq:epsilon_2}.

For a more general case, it can be useful to translate these coordinate-specific coefficients to a more general geometry. Writing the polarization state vector as $\mathbf{P}$ in terms of its components 
$\boldsymbol{\varepsilon_1}$ and $\boldsymbol{\varepsilon_2}$,
\begin{align}
    \mathbf{P} &= c_1 \boldsymbol{\varepsilon_1} + c_2 \boldsymbol{\varepsilon_2} 
    = c_1(-\sin\phi \mathbf{e}_x + \cos\phi \mathbf{e}_y)+ c_2 (\cos\theta \cos\phi \mathbf{e}_x + \cos \theta \sin \phi \mathbf{e}_y - \sin\theta \mathbf{e}_z) = \notag \\
    &= (-c_1 \sin\phi + c_2 \cos\theta \cos\phi) \mathbf{e}_x  +(c_1 \cos\phi + c_2 \cos \phi \sin\phi) \mathbf{e}_y - c_2 \sin \theta \mathbf{e}_z,
    \label{eq:polar_vec_cartesian}
\end{align}
allows us to instead express the polarization state vector $\mathbf{P}$ in terms of the Cartesian components with related coefficients
\begin{align}
    c_x &= -c_1 \sin \phi + c_2 \cos\theta \cos \phi \\
    c_y &= \phantom{+} c_1 \cos \phi + c_2 \cos \theta \sin \phi \\
    c_z &= -c_2 \sin \theta.
    \label{eq:polar_coefficients_cartesian}
\end{align}

We can now construct an observer-specific density matrix, relative to the three Cartesian unit vectors,
\begin{equation}
    \rho_{\rm obs} = \mathbf{P} \mathbf{P}^\dagger = 
    \left( {
        \begin{array}{ccc}
            c_x c^*_x   &   c_x c^*_y   &   c_x c^*_z \\
            c_y c^*_x   &   c_y c^*_y   &   c_y c^*_z \\
            c_z c^*_x   &   c_z c^*_y   &   c_z c^*_z 
        \end{array} }
    \right)
    \label{eq:rho_observer}
\end{equation}
which can be expressed in terms of the coefficients $c_1$ and $c_2$,
\begin{align}
    |c_x^2| &= |c_1^2| \sin^2\phi \notag \\
    &\phantom{= } - (c_1 c_2^* + c_2 c_1^*) \sin\phi \cos\theta \cos\phi \notag \\
    &\phantom{= } + |c_2^2| \cos^2\theta \cos^2\phi ,
    \label{eq:c_x^2} \\
    |c_y^2| &= |c_1^2| \cos^2\phi \notag \\
    &\phantom{=  } +  (c_1 c_2^* + c_2 c_1^*) \cos\phi \cos\theta \sin\phi \notag \\
    &\phantom{= } + |c_2^2| \cos^2\theta \sin^2\phi ,
    \label{eq:c_y^2} \\
    |c_z^2| &= |c_2^2| \sin^2\theta ,
    \label{eq:c_z^2} \\
    c_x c_y^*  &= -|c_1^2| \sin\phi \cos\phi \notag \\
                &\phantom{= } -c_1 c_2^* \sin\phi \cos\theta \sin\phi \notag \\
                &\phantom{= } +c_2 c_1^* \cos\theta\cos^2\phi \notag \\
                &\phantom{= } +|c_2^2| \cos^2\theta \cos\phi\sin\phi ,
    \label{eq:cx_cy} \\
    c_x c_z^*  &= c_1 c_2^* \sin\phi \sin\theta \notag \\
                &\phantom{= } - c_2 \cos\theta \cos\phi \sin \theta ,
    \label{eq:cx_cz} \\
    c_y c_z^*  &=  -c_1 c_2^* \cos\phi \sin\theta \notag \\
                &\phantom{= } -|c_2^2| \cos\theta \sin\phi \sin\theta ,
    \label{eq:cy_cz} \\
    c_x^* c_y  &=  (c_x c_y^*)^* ,\,
    c_x^* c_z  =  (c_x c_z^*)^* ,\,
    c_y^* c_z  =  (c_y c_z^*)^*,  \label{eq:cx_cy_cz_conj}
\end{align}
that can be retrieved directly from the original density matrix' components,
\begin{equation}
    |c_1^2| = \rho_{11},\, c_1 c_2^* = \rho_{12}, \, c_1^* c_2 = \rho_{21},\, |c_2^2| = \rho_{22},
    \label{eq:rho_cs}
\end{equation}
where $\rho_{12} = \rho_{21}$ when there is no circular polarization ($V=0$, and hence all elements are real).

\section{Comparisons}
\label{app:comparisons}
\subsection{Semi-Infinite Slab}
\label{app:slab}

\begin{figure}[htb]
    \centering
    \includegraphics[width=0.5\columnwidth]{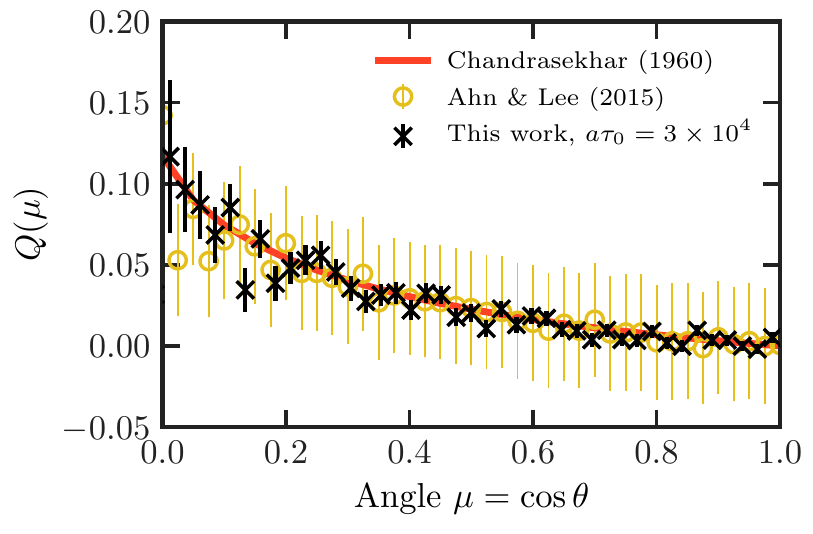}
    \caption{Degree of $Q$-polarization calculated in a semi-infinite plane-parallel slab of neutral hydrogen with $T=10$ K and $\tau_0 = 2\times10^6$ and the similar calculations by \cite{AHN2015POLARIZATIONHYDROGEN} who introduced the implemented density matrix formalism and an analytical solution by \cite{Chandrasekhar1960} for electron scatterings in an optically thick slab, showing good agreement.}
    \label{fig:slab}
\end{figure}

To test our implementation of the density matrix formalism in \texttt{tlac}, we compare it under the same geometry that was applied by \cite{AHN2015POLARIZATIONHYDROGEN}.
In Figure~\ref{fig:slab} we show results for a slab that is infinitely large in the $x,y$ directions.
This orientation ensures it to be aligned with the basis chosen to represent the density matrix.
The observable Stokes parameters do in this case coincide with the per-photon Stokes parameters.
These can be obtained from the density matrix components, with $I = \rho_{11} + \rho_{22}$, $Q=\rho_{11}-\rho_{22}$ and $U=\rho_{12} + \rho_{21}$.
We hence do not have to translate the components as in Appendix~\ref{app:translating_to_observer_basis}.
For the semi-infinite slab, polarization may only develop either parallel to the slab normal or perpendicular to it, which corresponds to the two basis vectors.
No cross-terms do then exists, and $U$ is zero.
In the figure, we then only plot $Q$ as a function of viewing angle $\mu = \cos \theta$ to the surface, where $\mu = 1$ corresponds to viewing the slab nadir.
The optical depth from the midplane of the slab is $2 \times 10^{6}$, and $T=10$ K.
This corresponds to the optically thick limiting case in which the photons escape from the damping wing after having undergone a diffusion process.
There, the polarization is given by the Rayleigh phase function, in which the polarization develops similarly to a process where a continuum photon is scattered by a free electron, for which \cite{Chandrasekhar1960} developed a solution, overplotted in the figure.
Our results fit those of \cite{AHN2015POLARIZATIONHYDROGEN} and \cite{Chandrasekhar1960} well, with larger polarization developing for sharper viewing angles ($\mu \to 0$).

\subsection{Universe Before Reionization: Scattering Out of Resonance Due to the Hubble Expansion}
\label{app:hubble}
We compare our polarization implementation with that of \cite{Rybicki1999}, who \citep[also through the companion paper][]{Loeb1999} examined scattering of Ly$\alpha$ photons through neutral hydrogen in an expanding Hubble volume, representative of the high-$z$, pre-reionization Universe. They predicted that Ly$\alpha$ is detectable as halos with a characteristic radius of $\sim 10-20$ arcseconds, and redshifted by several $10^3$ km s$^{-1}$ with respect to systemic due to the Hubble flow (the precise numbers depend on cosmological parameters). \cite{Rybicki1999} computed the polarization properties of these Ly$\alpha$ halos, and found them to be highly polarized, with the degree of linear polarization reaching $\sim60$\%. \cite{Rybicki1999} compared their results to \cite{Schuster1879}, where the polarization was directly related to the radial dependence of the product of the cross section and the column density. We, however, note that the degree of polarization is consistent with photons that have developed 100\% polarization in an early wing scattering that then is core-scattered forward (see Table~\ref{tab:QM_polarization}). 

    To test our polarization implementation, we repeated the analysis of \cite{Rybicki1999}, and compared results. We adopted $\Omega_{\rm b}=0.06$, $\Omega_{\rm m}=0.7$ and $\Omega_{\Lambda}=0.7$ and $h_0=0.7$, which translates to a mean present-day hydrogen number density $n_{\abbm{H}{1},0} = x_{\abbm{H}{1}} X n_{\rm b}$ where $n_{\rm b} =  2.5 \times 10^{-7}$ cm$^{-3}$ \citep{PlanckCollaborationXVI2014}, $x_{\abbm{H}{1}}$ is the neutral hydrogen fraction, and $X$ is the primordial number density of hydrogen, giving $n_{\abbm{H}{1}}(z) = n_{\abbm{H}{1},0}\left( 1+z \right)^3$. Following \cite{Loeb1999} and \cite{Rybicki1999} we assumed that $x_{\abbm{H}{1}} = 1$, and focused on $z=10$. We emitted $10^7$ photons centered at the Ly$\alpha$ line frequency, and set the core-skipping parameter $x_{\rm crit}=10$. We captured all photons escaping within 15$^\circ$ of the six principal axes, and binned the photons into logarithmic bins of their impact parameters. We then calculate the collective degree of linear polarization in each bin, $P(\alpha_x, \alpha_y)$ where $\alpha_x$ and $\alpha_y$ are their two dimensional impact parameters, as well as the variance. We calculate the uncertainty following the procedure outlined in App.~(\ref{app:uncertainty}). We then computed photon weighted averages of the linear polarization, $P(\alpha)$, where $\alpha$ now is the radial impact parameter.

Figure~\ref{fig:polardegree_tests}a compares our results to those of \cite{Rybicki1999}. We plot our normalized intensity $I/I_{\rm max}$ and the degree of total linear polarization $P(\alpha)$ as function of radial impact parameter $\alpha$ ({\it open black circles} with error bars), and compare to results from Figure~1 in \cite{Rybicki1999}, and rescale their normalized impact parameter $\tilde{p}$ using Eq.~(9) in \cite{Loeb1999} with our choice of cosmological parameters. We find our results to fit the rescaled results of \cite{Rybicki1999} well. We slightly underestimate $P(\alpha)$ at larger $\alpha$. We attribute the discrepancies in the degree of polarization to the choice of cosmology.

\subsection{Scattering in an Ionized Medium off an Expanding, Neutral Shell of Hydrogen}
\label{app:expanding_shell}
\begin{figure*}[htb]
    \centering
    \includegraphics[width=0.48\columnwidth]{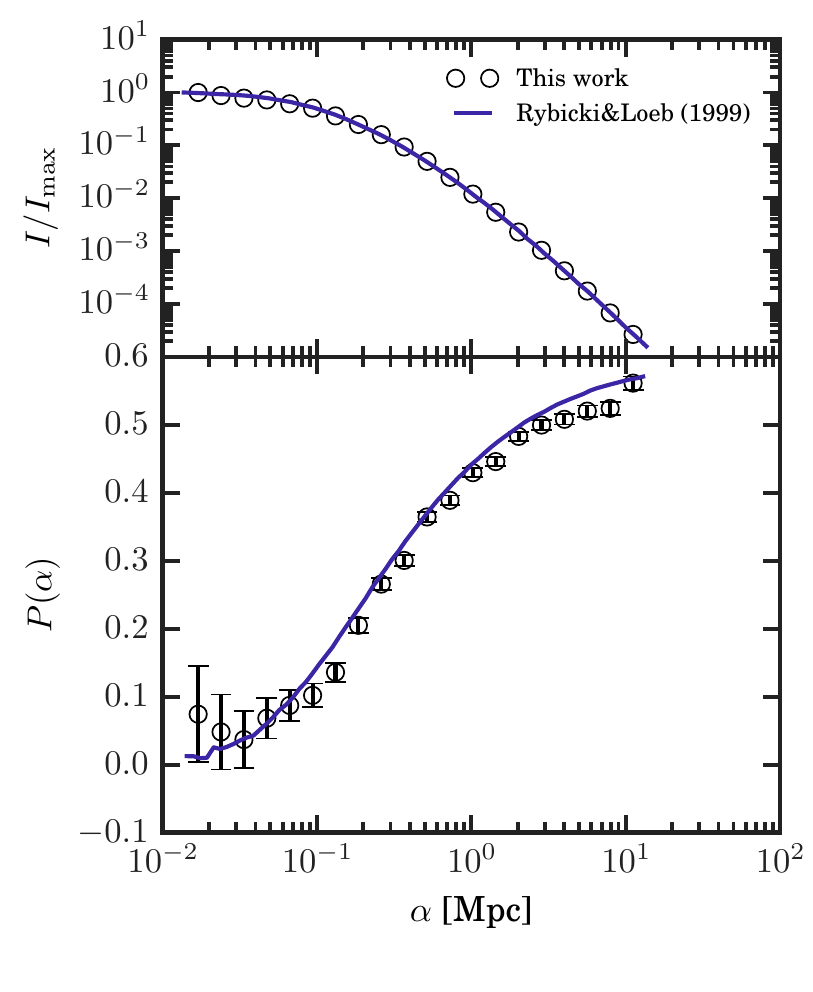}
    \includegraphics[width=0.48\columnwidth]{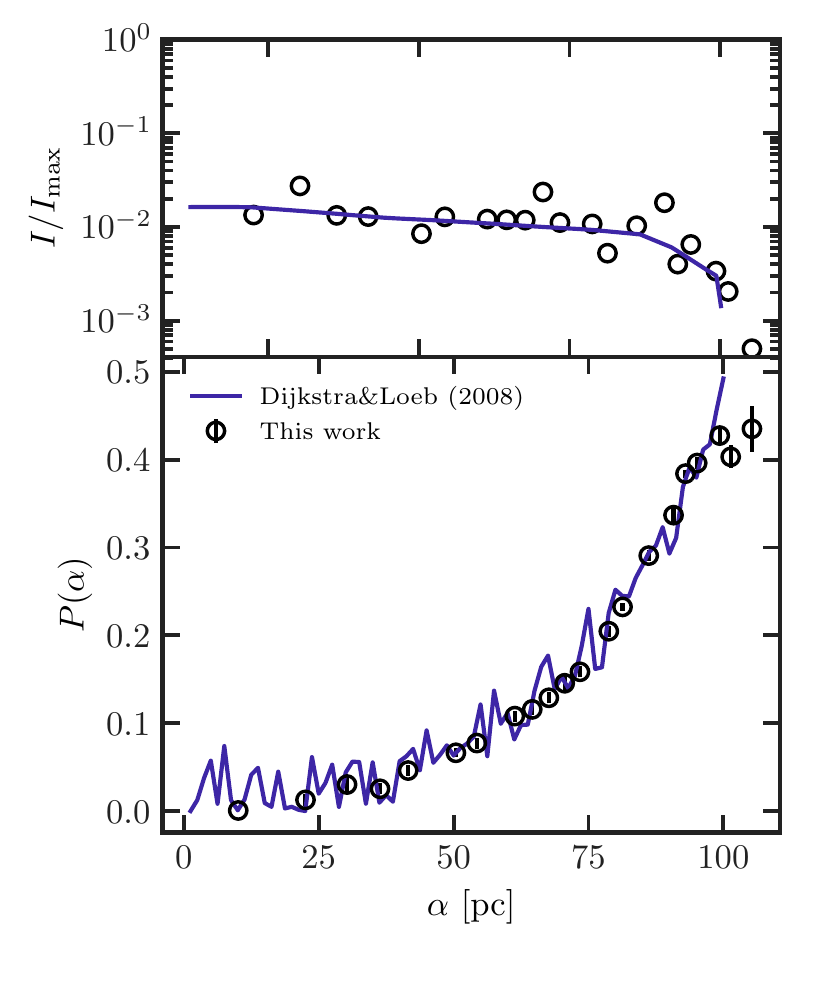}
    \caption{Upper panels: normalized intensity as function of impact parameter.
    Lower panels: degree of linear polarization against impact parameters.
    In panels (a), we compare our methods against the results of \cite{Rybicki1999} (solid, purple lines) for scattering in an Hubble-expanding medium of \abbm{H}{1} at $z=10$, and in panels (b), we compare our methods (empty, black circles, with error bars for $P(\alpha)$) against the results for an expanding shell of \abbm{H}{1} by \cite{Dijkstra2008} (solid, purple lines). }
    \label{fig:polardegree_tests}
\end{figure*}

    \cite{Dijkstra2008} explored the polarization of Ly$\alpha$ photons scattering through the often-used shell models. We repeat their analysis for a shell with column density $N_{\abbm{H}{1}}=10^{19}$~cm$^{-2}$ at $T=10^4$~K, $v_{\rm outflow} = 200$~km/s. In this simulation, we include all core scatterings (i.e. $x_{\rm crit}=0$). The central Ly$\alpha$ source emits unpolarized photons with frequencies Gaussian-distributed around the Ly$\alpha$ line center, with $\sigma=v_{\rm circ}$. Figure~\ref{fig:polardegree_tests}b compares the rescaled and then normalized intensity $I/I_{\rm max}$ and the degree of linear polarization $P(\alpha)$. Our results are represented by the {\it open black circles}, whereas those of \citet{Dijkstra2008} are represented by the {\it solid, purple line}. Both results agree well.

\section{Estimating uncertainty}
\label{app:uncertainty}

The polarization in each bin (pixel) is estimated from the contributions from all the photons falling into it. The arrival of photons on a detector is a Poissonian process, which for large number of photons is well described by Gaussian statistics. The variance can be estimated from the accumulated (or mean) polarization $P_j$ in a pixel $j$ and the per-photon polarization $P_{i,j}$,
\begin{equation}
    \sigma_j^2 = \frac{1}{N_{j}-1} \sum\limits_{i=1}^{N_j} \left( P_{i,j} - P_j  \right)^2.
    \label{eq:variance}
\end{equation}
To calculate the polarization as a function of the radial impact parameter $\alpha$, we bin the polarization in multiple pixels with (approximately) the same $\alpha$, 
\begin{equation}
    P(\alpha) = \frac{\sum_{j\in \alpha} w_j P_j}{\sum_{j\in \alpha} w_j }
    \label{eq:polarization_weighted}
\end{equation}
where we have weighted the contributions by the number of photons going into each pixel $j$, $N_j$, and the variance $\sigma_j^2$,
\begin{equation}
    w_j = \frac{N_j}{\sigma_j^2}.
    \label{eq:polarization_weigths}
\end{equation}
We calculate the standard error of the mean polarization $P(\alpha)$ by propagating the per-pixel standard error,
\begin{equation}
    \text{SE}\left(P(\alpha)\right) = \sqrt{\sum\limits_{j\in \alpha} \frac{1}{N_j}\left( \frac{\partial P(\alpha)}{\partial w_j}\sigma_j\right)^2 } = \frac{1}{\sum_{j\in \alpha} w_j} \sqrt{\sum\limits_{j\in \alpha} \frac{N_j}{\sigma_j^2}}.
    \label{eq:sem}
\end{equation}

\end{document}